\documentclass[times,sort&compress,10pt,3p]{elsarticle}

\usepackage[labelfont=bf]{caption}

\usepackage{amsmath,amsfonts,amssymb,amsthm,booktabs,color,epsfig,graphicx,hyperref,url}
\usepackage[linesnumbered]{algorithm2e}

\usepackage{cleveref}

\usepackage{xurl}

\usepackage{lipsum}
\makeatletter
\def\ps@pprintTitle{%
 \let\@oddhead\@empty
 \let\@evenhead\@empty
 \def\@oddfoot{}%
 \let\@evenfoot\@oddfoot}
\makeatother

\usepackage{chngcntr}
\theoremstyle{definition}
\newtheorem{theorem}{Theorem}

\newtheorem{remark}{Remark}

\newtheorem{lemma}{Lemma}

\theoremstyle{definition}

\newtheorem{condition}{Condition}

\usepackage{algpseudocode}
\usepackage{subcaption}

\DeclareMathOperator*{\argmax}{arg\,max}

\newcommand*{\bs}{\boldsymbol}
\newcommand*{\mb}{\mathbf}

\begin{document}

\begin{frontmatter}

\title{Projection-pursuit Bayesian regression for symmetric matrix predictors}

\author[]{Xiaomeng Ju \corref{mycorrespondingauthor}}
\author[]{Hyung G. Park}
\author[]{Thaddeus Tarpey}
\address{Division of Biostatistics, Department of Population Health \\ 
New York University School of Medicine, New York, NY, United States}

\cortext[mycorrespondingauthor]{Email address: \url{jux01@nyu.edu}}

\begin{abstract}
This paper develops a novel Bayesian approach for nonlinear regression with  symmetric matrix predictors, often used to encode connectivity of different nodes.   Unlike methods that vectorize matrices as predictors that result in a large number of model parameters and unstable estimation, we propose a Bayesian multi-index regression method, resulting in  a projection-pursuit-type estimator that leverages the structure of matrix-valued predictors.  We establish the model  identifiability conditions and impose a sparsity-inducing prior on the projection directions for sparse sampling to prevent overfitting and enhance interpretability of the parameter estimates.  Posterior inference is conducted through Bayesian backfitting.  The performance of the proposed method is evaluated through simulation studies and a case study investigating the relationship between brain connectivity features and cognitive scores.  
\end{abstract}


\end{frontmatter}

\section{Introduction\label{sec:1}}
In many applications, matrices, instead of real-valued vectors, are needed for prediction in regression models. These predictors often encode the connections between individual features, such as brain functional connectivity \citep{bastos2016tutorial,weaver2023single}, gene co-expression networks \citep{zhang2005general,langfelder2008wgcna}, and social networks \citep{abbasi2011identifying,zhang2020logistic}, in the form of symmetric matrices.   In this paper, we  develop a flexible regression method that accounts for the symmetric matrix structure in the predictors, enabling accurate prediction and interpretation of the nonlinear relationships between matrices and the response.  

A major challenge with matrix-variate regression is the high dimensionality of the predictor space, as a $p \times p$ matrix contains $O(p^2)$ elements.   Classical regression methods can be applied  using vectorized matrices as predictors, often resulting in high dimensional predictor spaces that require variable selection or dimension reduction.  They may not yield strong and interpretable features for the outcome and can lead to overfitting due to overparameterization.

To address these issues, various methods have been developed specifically for matrix or higher-order tensor predictors, taking into account their high-dimensionality, with the majority of these methods assuming a linear model. For matrix predictors  $\mb{M}_i \in \mathbb{R}^{p \times q}$ and response 
 $Y_i \in \mathbb{R}$, the model is given by   $E(Y_i|\mb{M}_i) = \langle \mb{M}_i, \mb{C} \rangle$ for $i \in  \{1, \dots , n\}$, with  $\langle \mb{M}_i, \mb{C} \rangle =  \text{vec}(\mb{M}_i)^T \text{vec}(\mb{C})$ being the Frobenius inner product and  $\mb{C} \in \mathbb{R}^{p \times q}$ being the coefficient matrix to be estimated.   To avoid overfitting,  regularization on $\mb{C}$ has been considered by adding penalties in frequentist methods \citep{zhou2014regularized, reiss2015wavelet, wang2017generalized, raskutti2019convex, chen2019non, zhou2021improved,brzyski2023matrix} or applying informative priors in Bayesian approaches  \citep{goldsmith2014smooth,kang2018scalar,guha2021bayesian}. These techniques encourage desired properties of $\mb{C}$, such as sparsity, smoothness, and low-rank.  In particular, for symmetric matrix predictors, \citet{guha2021bayesian}  and \citet{brzyski2023matrix} consider a symmetric coefficient $\mb{C}$ and impose row/column wise,  entry-wise, and spectral regularization on  $\mb{C}$. Another popular modeling choice is through a tensor decomposition  \citep{kolda2009tensor} of the coefficient  $\mb{C}$, which includes proposals based on a rank-1 decomposition \citep{guo2011tensor,zhao2014structured}, the CANDECOMP/PARAFAC (CP) decomposition \citep{zhou2013tensor, guhaniyogi2017bayesian, he2018boosted, papadogeorgou2021soft}, and the Tucker decomposition   \citep{li2018tucker,zhang2020islet}. In particular, for 
   symmetric matrices, a symmetric bilinear decomposition has been considered by \citet{wang2019symmetric} and \citet{ju2024bayesian}.  Many of the  linear model-based proposals also apply to generalized linear settings for $Y_i$'s that follow exponential families of distributions, 
   see e.g. \cite{zhou2013tensor,zhou2014regularized,wang2017generalized,guhaniyogi2017bayesian,li2018tucker,chen2019non,papadogeorgou2021soft}. However, the (generalized) linear assumptions on the regression function can be restrictive to approximate complex relationships often encountered in practice.

 Beyond linear models,  nonlinear  strategies have been explored under additional assumptions on the tensor predictors. 
  Several authors used Gaussian processes (GP) to model nonlinear relationships, where  the kernel or mean functions defining the GP are based on a generative model of the tensor predictors \citep{zhao2014tensor,hou2015online,ma2022semi}.   A few nonlinear regression methods assume 
a low-rank decomposition of the tensor predictors    
 $\mb{X}_i \in \mathbb{R}^{I_1 \times I_2 \times ...\times I_K}$. For a rank-1 decomposition $\mb{X}_i = \mb{x}_{1,i} \circ \cdots \circ \mb{x}_{K,i}$,  where $\circ$ denotes the outer product and $\mb{x}_{k,i} \in \mathbb{R}^{I_k}$ for $k \in  \{1,..., K\}$, \citet{signoretto2013learning} considered
 a  model of the form $E(Y_i|\mb{X}_i) = \sum_{r=1}^R \prod_{k=1}^Kf_{(r,k)}(\mb{x}_{k,i})$,  where the $I_k$-variate functions $f_{(r,k)}$'s in  reproducing kernel Hilbert spaces (RKHS) can be estimated from the representer theorem. Alternative estimation algorithms and extensions to the CP decomposition have been studied by several authors \citep{kanagawa2016gaussian,suzuki2016minimax,imaizumi2016doubly}. By setting $K = 2$ and $I_1 = I_2 = p$, these methods can be adapted to regression with symmetric matrix predictors. However, the assumed decomposition structure of $\mb{X}_i$ may be misspecified in practice.   Moreover, estimating $p$-variate nonparametric functions ($f_{(r,k)}$'s) can be challenging when $p$ is large, and the resulting estimates can be difficult to interpret since they are based on the decomposed components ($\mb{x}_{k,i}$) of the matrix predictors.

On the other hand, nonlinear additive models 
 for tensor regression rely on univariate functions defined on the entries of the tensor predictors. 
For matrices $\mb{M}_i \in \mathbb{R}^{p\times q}$, the model is given by $E(Y_i|\mb{M}_i)  = \sum_{j = 1}^p \sum_{k = 1}^q f_{j,k}\left(\mb{M}_{i,(j,k)}\right)$, where $\mb{M}_{i, (j,k)}$ is the $(j,k)$-th element of $\mb{M}_i$, and each $f_{j,k}: \mathbb{R} \rightarrow \mathbb{R}$ is a smooth function. To estimate the functions ($f_{j,k}'s$), \citet{hao2021sparse} and \citet{chen2023bayesian} suggested spline-based methods: the former using a frequentist approach with low-rank and sparse penalties, and the latter adopting a Bayesian framework with sparsity-inducing priors.  While allowing for interpreting the contribution of each element of $\mb{M}_i$ , these methods face challenges in accurately estimating many nonparametric functions and can lead to overfitting when the regression function is not very sparse across the elements of $\mb{M}_i$. \citet{zhou2024broadcasted} introduced an alternative additive model that transforms each element of $\mb{M}_i$ using the same  functions  $f_r: \mathbb{R} \rightarrow \mathbb{R}$, for $r \in  \{1,..., R\}$, and constructs a regression model $E(Y_i|\mb{M}_i)  = \sum_{r = 1}^R \langle \mb{B}_r,  \mb{F}_{i,r} \rangle$, with the $(j,k)-$th element of the transformation $\mb{F}_{i,r, (j,k)} = f_r\left(\mb{M}_{i, (j,k)}\right)$ and the coefficients $\mb{B}_r \in \mathbb{R}^{p \times q}$. This model captures the nonlinearity through $f_r$'s, but it restrictively assumes that identical transformations of all elements of $\mb{M}_i$ will yield highly predictive features.

Supervised dimension reduction offers another solution to the high-dimensionality of  predictors by projecting them to a lower-dimensional space that best captures the predictive information.  This approach has not been widely explored in the context of matrix-variate regression. \citet{ding2015tensor} generalized the Sliced Inverse Regression (SIR) \citep{li1991sliced} to matrix predictors, assuming that
 each column/row of $\mb{M}_i$ is a linear combination of its columns/rows. 
 \citet{marx2011multidimensional} and \citet{weaver2023single} considered a single-index model $E(Y_i|\mb{M}_i) = g(\langle \mb{M}_i, \mb{C} \rangle)$  with the Frobenius inner product $\langle \cdot, \cdot \rangle$ and an unknown link function $g$. Smoothness and sparsity assumptions are imposed on $\mb{C}$, resulting in a parsimonious estimator that captures nonlinear relationships while allowing interpreting entries of $\mathbf{C}$ as contributions from corresponding entries in $\mathbf{M}_i$. We adopt this single-index structure as  the building block of our proposal.


We propose a Bayesian multi-index regression method  for symmetric matrix predictors, resulting in a projection-pursuit-type estimator.  This approach serves as a supervised dimension reduction tool, taking the symmetry of matrices into account in finding the projections that contain the most predictive information for the response. 
  To avoid overfitting and to adapt to cases where only a subset of matrix entries are related to the response, a spike-and-slab \citep{rovckova2018spike} prior is used to perform sparse sampling of the projection directions.  Unlike existing frequentist  proposals, our Bayesian method provides uncertainty quantification of the regression estimates.   Furthermore, the proposed approach does not rely on restrictive assumptions of generative models or tensor decompositions of the matrix predictors, as required by many other proposals (e.g. \cite{signoretto2013learning,zhao2014tensor,hou2015online,kanagawa2016gaussian,suzuki2016minimax,ding2015tensor,imaizumi2016doubly,ma2022semi}).  A recent work of \citet{collins2024bayesian} studied Bayesian projection pursuit regression for vector predictors. Expanding from their proposal, we use single-index estimators for symmetric matrix predictors as the additive components and adopt a sparsity-inducing prior that involves fewer parameters. Our method can be used as a semi-parametric approach for inference and prediction, or as a nonparametric approach for prediction only.   For inference, it allows for the interpretation of projection directions and identifies the contribution of each predictor matrix entry to the response.  For prediction, it provides a flexible tool to approximate nonlinear regression functions, with the number and the complexity of single-index estimators regulating the overall complexity of the fit.

  The remainder of this article is organized as follows. \Cref{sec:methodology} introduces the regression model and establishes its identifiability conditions. This is followed by a description of posterior inference 
  through Bayesian backfitfing and  sparse Markov chain Monte Carlo (MCMC) sampling from single-index models based on partial residuals. \Cref{sec:simulation} presents simulation experiments evaluating the performance of our proposal in comparison to existing alternatives.  
     \Cref{sec:hcp} describes an application to data from the Human Connectome Project (HCP), using brain functional connectivity to predict cognitive scores. Finally, \Cref{sec:discussion} concludes with a discussion of our approach and results.

\section{Methodology} \label{sec:methodology}
In this section, we develop a regression model for a scalar outcome using symmetric matrix predictors,  followed by introducing the posterior inference procedure. 

\subsection{Regression model}
Consider $p \times p$ symmetric matrix predictors $\mb{M}_i \in \text{Sym}^{p}$ and responses $Y_i \in \mathbb{R}$, for $i \in  \{1,..., n\}$. We define a regression model as follows
\begin{align} \label{eq:model}
Y_i &=  \mu + \sum_{k=1}^K g_k \left(\left \langle \mb{M}_i, \mb{C}_k \right \rangle \right) + \epsilon_i,  \ \epsilon_i \sim N(0, \sigma^2),
\end{align}
where  $\mu \in \mathbb{R}$,  $\epsilon_i$ is independent from $\mb{M}_i$, $\langle\  , \rangle $ denotes the Frobenius inner-product, and each $g_k$ is an unknown function called the ``ridge function"  \citep{friedman1981projection},  with the corresponding coefficient matrix $\mb{C}_k  = \bs{\gamma}_k\bs{\gamma}_k^T \in \mathbb{R}^{p \times p}$ constructed with projection direction $\bs{\gamma}_k \in \mathbb{R}^p$ that satisfies the Euclidean L2 norm constraint $||\bs{\gamma}_k||_2 = 1$.  The proposed model~\eqref{eq:model} is additive in single index terms, where each additive term $g_k \left( \langle \mb{M}_i,  \mb{C}_k \rangle \right)$ is computed through a projection of  the matrix predictor $\mb{M}_i$ onto the direction  $\bs{\gamma}_k$.  This formulation makes each additive term $g_K(\langle \mb{M}_i, \mb{C}_k \rangle )$ correspond to a special case of the matrix single-index model studied by \citet{marx2011multidimensional} and \citet{weaver2023single}, which assumes $E(Y_i|\mb{M}_i) = g(\langle \mb{M}_i, \mb{C} \rangle)$ for a more generic coefficient $\mb{C} \in \mathbb{R}^{p \times p}$. In model \eqref{eq:model}, the symmetric coefficient matrix $\mb{C}_k$ is assumed to be  rank-1, yielding a parsimonious estimator for each term, and the additive combination of these individual terms enhances the model's flexibility.

\subsection{Identifiability} \label{subsection:identifiability}
Additive index models provide greater flexibility compared to linear and single-index models, but without identifiability, they can be difficult to interpret, thus limiting their use as dimension reduction tools. In multivariate settings, the identifiability of additive index models has been studied by several authors \cite{chiou2004quasi,lin2007identifiability,yuan2011identifiability}. Among them, \citet{yuan2011identifiability} imposed the most relaxed assumptions on the ridge functions. We extend their results to our model \eqref{eq:model} and show its identifiability under mild conditions introduced below.

\begin{condition}
 \label{assumption:1}
	 A set of continuous univariate functions $ \mathcal{F} = \{f_1, \dots, f_d\}$ and  projection directions $\mb{B} = (\bs{\beta}_1, \dots, \bs{\beta}_d) \in \mathbb{R}^{p \times d}$, $d < p$,  such that $\bs{\beta}_j \in \mathbb{R}^p$ and $||\bs{\beta}_j||_2 = 1$  for $j = 1, \dots ,d$,   satisfy  
\begin{itemize}
\item[(a)] $f_j(0) = 0$ and $f_j \not\equiv 0$ for $j \in \{1, \dots, d\}$; 
\item[(b)] there is at most one linear function in $\mathcal{F}$; 
\item[(c)]  $\mb{B}$ and  $\mb{B} \odot \mb{B}$ are column full-rank, where $\odot$ denotes element-wise multiplication (i.e. the Hadamard product)
\end{itemize}
\end{condition}

\begin{theorem} \label{theorem:1}
Suppose that $\mu, \tilde{\mu} \in \mathbb{R}$, and \Cref{assumption:1} holds for $\{\mathcal{G}, \bs{\Gamma}\}$ and $\{\tilde{\mathcal{G}}, \tilde{\bs{\Gamma}}\}$ respectively, where $\mathcal{G} = \{g_1,\dots, g_K\}$, $ \bs{\Gamma} = (\bs{\gamma}_1,\dots, \bs{\gamma}_K) \in \mathbb{R}^{p \times K}$ ,  $\tilde{\mathcal{G}} = \{\tilde{g}_1,\dots, \tilde{g}_{\tilde{K}}\}$,  and $\tilde{\bs{\Gamma}} = (\tilde{\bs{\gamma}}_1,\dots, \tilde{\bs{\gamma}}_{\tilde{K}}) \in \mathbb{R}^{p \times \tilde{K}}$,  and if 
\begin{equation}	
\label{eq:identify}
\mu + g_1(\bs{\gamma}_1^T\mb{M} \bs{\gamma}_1) + \cdots + g_K(\bs{\gamma}_K^T\mb{M} \bs{\gamma}_K) = \tilde{\mu} + \tilde{g}_1(\tilde{\bs{\gamma}}_1^T\mb{M} \tilde{\bs{\gamma}}_1) + \cdots + \tilde{g}_{\tilde{K}}(\tilde{\bs{\gamma}}_{\tilde{K}}^T\mb{M} \tilde{\bs{\gamma}}_{\tilde{K}}),
\end{equation}
for  any matrix $\mb{M} \in \text{Sym}^p$, then the additive index model \eqref{eq:model} is identifiable in the following sense:
\begin{itemize}
\item[(a)] the intercepts agree: $\mu = \tilde{\mu}$;
\item[(b)] the dimensionalities agree: $K = \tilde{K}$;
\item[(c)] there exist a permutation $\pi(1), \dots, \pi(K)$  of $\{1, \dots, K\}$, and $l_j \in \{0, 1\}$ for $j \in \{1,\dots, K\}$,   such that 
$$\bs{\gamma}_j = (-1)^{l_j} \tilde{\bs{\gamma}}_{\pi(j)}, \ g_j = \tilde{g}_{\pi(j)}.$$
\end{itemize}
\end{theorem}

The proof of \Cref{theorem:1} is provided in Appendix A.

\begin{remark}
 \Cref{theorem:1} establishes the identifiability of the proposed model for  $\mb{M} \in \text{Sym}^p$. In fact, the identifiability also holds for $p \times p$ symmetric positive definite (SPD) matrices $\mb{M} \in \text{SPD}^{p}$ under the same assumptions.  This is particularly useful in various applications where the matrix predictors are SPD, such as covariance or correlation matrices, as our model can be applied while maintaining its interpretability.  
\end{remark}

Note that the requirements in  \Cref{assumption:1} placed on the ridge functions are more relaxed compared to those imposed in \citet{yuan2011identifiability} for the original projection pursuit regression to be identifiable. \citet{yuan2011identifiability}'s   conditions restrict the number of linear or quadratic ridge functions to no more than one, while we only limit the number of linear ridge functions to one. Below, we illustrate why restricting the model to have at most  one linear ridge function is necessary for identifiability.  
\begin{remark}
Consider two distinct linear functions $g_1(u) = a_1 u$ and $g_2(u) = a_2 u$, along with projection directions $\bs{\gamma}_1$ and $\bs{\gamma}_2$, where $||\bs{\gamma}_1||_2 = ||\bs{\gamma}_2||_2= 1$.  Let $\mb{D} = a_1\bs{\gamma}_1 \bs{\gamma}_1^T+ a_2\bs{\gamma}_2 \bs{\gamma}_2^T$ which can be represented by its eigendecomposition $\mb{D} = \lambda_1\mb{v}_1\mb{v}_1^T + \lambda_2\mb{v}_2\mb{v}_2^T$, where $\lambda_1$ and $\lambda_2$ are the eigenvalues, and  $\mb{v}_1$ and $\mb{v}_2$ are the eigenvectors satisfying $||\mb{v}_1||_2 = ||\mb{v}_2||_2 = 1$ and $\mb{v}_1^T\mb{v}_2 = 0$.  For any $\mb{M} \in \text{Sym}^p$, it can be shown that
$$g_1(\bs{\gamma}_1^T\mb{M}\bs{\gamma}_1) + g_2(\bs{\gamma}_2^T\mb{M}\bs{\gamma}_2) = \langle \mb{M}, a_1\bs{\gamma}_1 \bs{\gamma}_1^T+ a_2\bs{\gamma}_2 \bs{\gamma}_2^T \rangle = \langle \mb{M}, \lambda_1\mb{v}_1\mb{v}_1^T + \lambda_2\mb{v}_2\mb{v}_2^T \rangle = \tilde{g}_1(\mb{v}_1^T\mb{M}\mb{v}_1) +  \tilde{g}_2(\mb{v}_2^T\mb{M}\mb{v}_2),$$
where $\tilde{g}_1(u) = \lambda_1 u$ and $\tilde{g}_2(u) = \lambda_2 u$. If $\bs{\gamma}_1^T \bs{\gamma}_2 \neq 0$, then $\{\bs{\gamma}_1, \bs{\gamma}_2\} \neq \{ \mb{v}_1, \mb{v}_2\}$, which results in a model that is unidentifiable.  
\end{remark}

For a model with multiple strictly linear ridge functions to be identifiable,  additional assumptions, such as the orthogonality of the projection directions, are required. However, we prefer to keep the model flexible  by not imposing these constraints. Based on our experience, our method  remains effective for prediction purposes  even if the true regression function contains multiple strictly linear functions.   

\begin{remark} \label{remark:3}
\Cref{assumption:1} (a) requires $g_k(0) = 0$ for $k \in  \{1,\dots, K\}$, which fixes the value of $g_k$ at index 0. In practice, if the indices are sparse around 0, this may lead to unstable estimates of $g_k$'s.  Instead, we adopt a revised condition: $g_k$ satisfies \Cref{assumption:1}, except for   $g_k(0) = 0$ being replaced by  $\sum_{i=1}^n g_k(\bs{\gamma}_k^T \mb{M}_i\bs{\gamma}_k) = 0$.  To see the equivalence of the original and revised conditions, let $h_k(u)= g_k(u) - g_k(0)$ for $u \in \mathbb{R}$, which ensures that $h_k(0) = 0$ and $h_k$ satisfies \Cref{assumption:1}.   \Cref{theorem:1} implies the identifiability of $h_k$'s in the model $E(Y_i|\mb{M}_i) =  \nu + \sum_{k=1}^K h_k \left(\bs{\gamma}_k^T \mb{M}_i \bs{\gamma}_k \right)$ for $\nu \in \mathbb{R}$, $h_k$'s and $\bs{\gamma}_k$'s that satisfy  \Cref{assumption:1}.   Since $g_k$ differs from $h_k$ by a constant $g_k(0)$ and that $g_k(0)$ can be determined  given $h_k$ due to $\sum_{i=1}^n g_k(\bs{\gamma}_k^T \mb{M}_i\bs{\gamma}_k) = 0$, the identifiability for $g_k$ holds for $k \in \{1,\dots, K\}.$
\end{remark}

\RestyleAlgo{ruled}

\subsection{Bayesian backfitting}
\label{subsec:backfitting}
For the proposed regression model \eqref{eq:model}, we introduce a posterior sampling framework  using Bayesian backfitting \citep{hastie2000bayesian}. This iterative procedure fits additive models by sampling each component from its conditional posterior distribution one at a time while keeping the other components fixed. It can be viewed as a form of Gibbs sampling~\citep{geman1984stochastic} where each additive component is treated as a parameter to be sampled.  The procedure is detailed in \Cref{alg:pbr}. 
 
 
 The algorithm begins by initializing $\mu$, $\sigma^2$, $g_{k}$,  and $\bs{\gamma}_{k}$, with initial values as $\mu_0$ , $\sigma_0$,  $g_{k,0}$,  and $\bs{\gamma}_{k,0}$ respectively, for $k \in  \{1,\dots, K\}$.  
While Bayesian backfitting does not strictly require good initial points and should converge to the posterior distribution given a sufficiently large number of iterations,  good initial points can accelerate convergence.   We set $\mu_0 = \sum_{i=1}^n Y_i/n$, $\sigma_0^2 = \sum_{i=1}^n (Y_i - \mu_0)^2 /(n-1)$ and $g_{k,0}  \equiv 0$ for $k \in \{1,\dots, K\}$, and set $\bs{\gamma}_{k,0}$'s based on an initial fit of the projection pursuit regression (PPR) \citep{friedman1981projection} detailed in \Cref{sec:simulation}. 
 
 Let  $\mathcal{D} = \{(\mb{M}_i, Y_i), i \in  \{1,\dots, n\}\}$ denote the observed data. 
\Cref{alg:pbr} adopts an MCMC approach to sample from the posterior distribution $p \left( \{g_k\}_{k=1}^K, \{\bs{\gamma}_k\}_{k=1}^K, \mu, \sigma^2 | \mathcal{D}\right)$, with each iteration $t \in  \{1,\dots, T\}$ consisting of three steps. Step~1 alternates over the $K$ additive components, sampling  the ridge function $g_k$ and projection direction $\bs{\gamma}_k$, and centering the sampled $g_k$ for identifiability (see \Cref{remark:3}). This step amounts to sampling from a single-index model based on partial residuals, as will be explained in \Cref{subsec:sim}.
 Step 2 samples the variance $\sigma^2$ conditioned on the current additive components and the  global mean  $\mu$ sampled from the previous iteration.  Step 3 updates the global mean $\mu$ based on the sampled additive components and $\sigma^2$. The use of conjugate priors for $\mu$ and $\sigma^2$ simplifies posterior sampling in Steps 2 and 3.  In the following, we provide more details on each step of \Cref{alg:pbr}.

\textbf{Step 1:} At the $t$-th iteration, to sample from the $k$-th additive term, we calculate the partial residuals $\mb{r}_{t-1}^{(k)} = \left(r_{1, t-1}^{(k)}, \dots, r_{n, t-1}^{(k)}\right)$ where 
$r_{i,t-1}^{(k)} = Y_i - \mu_{t-1} - \sum_{l \neq k} g_{l,t_l} \left( \bs{\gamma}_{l,t_l}^T \mb{M}_i \bs{\gamma}_{l,t_l} \right)$, for $i = 1,\dots, n,$ with $t_l = t$ for $l < k$ and  $t_l = t-1$ for $l \geq k$. Let $G_{t-1}^{(k)} =  \{ g_{l,t_l}\}_{l=1}^K$ and $\Gamma_{t-1}^{(k)} = \{\bs{\gamma}_{l, t_l}\}_{l=1}^K$ represent the most recently sampled $K$ ridge functions and $K$ projection directions.  Since  $p \left( g_k, \bs{\gamma}_k  \middle \vert  \ \mu_{t-1}, G_{t-1}^{(k)}, \Gamma_{t-1}^{(k)}, \mathcal{D} \right)$ can be simplified to $p \left( g_k, \bs{\gamma}_k  \middle \vert  g_{k, t-1},  \bs{\gamma}_{k, t-1}, \mathcal{R} \right)$, where $ \mathcal{R} = \left \{ 
\left(\mb{M}_i, r_{i, t-1}^{(k)} \right), i \in  \{1,\dots, n\} \right \}$, sampling  $g_k$ and $\bs{\gamma}_k$ is equivalent to sampling from a single-index model with responses $\mb{r}_{t-1}^{(k)}$. Further details on  the sampling procedure are provided in \Cref{subsec:sim}.

 \textbf{Step 2:}  Assuming the prior $\sigma^2 \sim IG(\alpha, \beta)$, the conditional posterior distribution of $\sigma^2$ is given by (see Appendix B for derivation): 
  $$\sigma^2 |\mu_{t-1}, \{g_{k,t}\}_{k=1}^K, \{\bs{\gamma}_{k,t}\}_{k=1}^K,  \mathcal{D} \sim  IG \left(\tilde{\alpha}, \tilde{\beta}\right),$$
 where $\tilde{\alpha} = \alpha + n/2$ and $\tilde{\beta} =  \beta + \sum_{i=1}^n \left(Y_i - \sum_{k=1}^Kg_{k,t}(\bs{\gamma}^T_{k,t}\mb{M}_i\bs{\gamma}_{k,t}) - \mu_{t-1}\right)^2/2$.

\textbf{Step 3:} Assuming the prior $\mu \sim N(a, b^2)$, the conditional posterior distribution of $\mu$ is given by   (see Appendix B for derivation): 
  $$\mu |\sigma^2_{t}, \{g_{k,t}\}_{k=1}^K, \{\bs{\gamma}_{k,t}\}_{k=1}^K,  \mathcal{D} \sim N \left(\tilde{\mu}, \tilde{\sigma}^2 \right),$$
where $\tilde{\sigma}^2 = \left( 1/b^2 + n/\sigma_{t}^2 \right)^{-1}$ and $\tilde{\mu} = \tilde{\sigma}^2 \left(a/b^2 + \sum_{i=1}^n \left(Y_i - \sum_{k=1}^Kg_{k,t}(\bs{\gamma}^T_{k,t}\mb{M}_i\bs{\gamma}_{k,t})\right)/\sigma_{t}^2\right).$  \\

At a higher level, \Cref{alg:pbr} performs Gibbs sampling \citep{geman1984stochastic} with two blocks: $B_1 = \left \{\{g_j\}_{j=1}^K,\{\bs{\gamma}_j\}_{j=1}^K, \sigma^2 \right\}$ (Steps 1 and 2), and $B_2 = \{\mu \}$ (Step 3). In particular, for the fist block, Step~1 samples from the conditional posterior distribution  marginalized over $\sigma^2$, $p \left(\{g_{k}\}_{k=1}^K, \{\bs{\gamma}_{k}\}_{k=1}^K \middle| \  \mu, \mathcal{D}\right)$. Step 2 samples $\sigma^2$, given the sampled $\{g_k\}_{k=1}^K$ and $\{\bs{\gamma}_k\}_{k=1}^K$, 
from  $p\left(\sigma^2 \middle | \ \{g_{k}\}_{k=1}^K, \{\bs{\gamma}_{k}\}_{k=1}^K, \mu, \mathcal{D}\right)$. Together, Steps 1 and 2 produce samples from $p(B_1|B_2, \mathcal{D})$. We iterate Steps~1-3 for a total of $T$ iterations, including warm-up and sampling  iterations. The samples obtained after $T'$ warm-up iterations are used for inference and prediction. \Cref{alg:pbr} embeds \Cref{alg:sim} that involves the posterior sampling from individual single-index models (see \Cref{subsec:sim}).

\SetKwProg{Input}{Input:}{}{}
\SetKwProg{Return}{Return:}{}{}
\SetKwProg{Initialization}{Initialize:}{}{}
\SetKwProg{Parameters}{Parameters:}{}{}
\SetKwFor{For}{For}{do}{EndFor}
\SetKwProg{DefData}{Data:}{}{}

\SetInd{0.3em}{3em}
\begin{algorithm}\caption{Projection-pursuit Bayesian regression}
\label{alg:pbr}

\DefData{$\mathcal{D} = \{ (\mb{M}_i, Y_i), i \in \{1,\dots, n\}\}$}{}
\Initialization{$\mu_0$ , $\sigma_0$,  $g_{k,0}$,  $\bs{\gamma}_{k,0}$,   for $k \in  \{1, \dots, K\}$ }{} 
\For{$t \in  \{1, \dots, T\}$}{

\hspace{-0.6cm}
\textbf{Step 1:} \\
\For{$k \in  \{1, \dots, K\}$}{
  $t_l = t$ for $l <k$, and $t_l = t-1$ for $l \geq k$ \\
  $G_{t-1}^{(k)} = \left \{ g_{l,t_l} \right\}_{l=1}^K$ and $\Gamma_{t-1}^{(k)} = \{\bs{\gamma}_{l, t_l} \}_{l=1}^K$ \\
 Sample $g_{k,t}$ and $\bs{\gamma}_{k,t}$ from $p \left( g_k, \bs{\gamma}_k  \middle \vert  \ \mu_{t-1},  G_{t-1}^{(k)}, \Gamma_{t-1}^{(k)}, \mathcal{D} \right)$  \algorithmiccomment{See \Cref{alg:sim}}\\
Center $g_{k,t} = g_{k,t} - \frac{1}{n}\sum_{i=1}^n g_{k,t} \left(\bs{\gamma}_{k,t}^T \mb{M}_i \bs{\gamma}_{k,t}\right)$  \\
}{}  
 \hspace{-0.6cm}
\textbf{Step 2:} \\
 Sample  $\sigma_t^2$ from $p \left(\sigma^2 \  \middle| \ \mu_{t-1}, \{g_{k, t} \}_{k=1}^K,  \{\bs{\gamma}_{k,t} \}_{k=1}^K,  \mathcal{D} \right)$\\
 \hspace{-0.6cm}
\textbf{Step 3:} \\
 Sample  $\mu_{t}$ from $p \left(\mu \ \middle| \ \sigma_{t}^2, \{g_{k, t} \}_{k=1}^K,  \{\bs{\gamma}_{k,t} \}_{k=1}^K,  \mathcal{D} \right)$ \\
 }
 
 \Return{$\mu_t$, $\sigma_t^2$, $g_{k,t}$, and $\bs{\gamma}_{k,t}$,  for $k \in  \{1,\dots, K\}$, $t \in  \{T',\dots, T\}$}{}
\end{algorithm}

\subsection{Sparse sampling for single index model} \label{subsec:sim}
Consider sampling from $p \left( g_k, \bs{\gamma}_k  \middle \vert  g_{k, t-1}, \bs{\gamma}_{k,t-1}, \mathcal{R} \right)$ in Step 1 of \Cref{alg:pbr}, which can be formulated as sampling from the posterior distribution derived from a single-index model 
\begin{equation} \label{eq:model:sim}
	r_i = g(\bs{\gamma}^T \mb{M}_i \bs{\gamma}) + \epsilon_i, \ i \in  \{1,\dots, n\},
\end{equation}
where $r_i = r_{i,t-1}^{(k)}$, $||\bs{\gamma}||_2 = 1$, and $\epsilon_i \sim N(0, \sigma^2)$. To account for the constraint $||\bs{\gamma}||_2 = 1$, we introduce a transformation $q: \mathbb{R}^{p-1} \rightarrow \mathbb{R}^{p}$ , mapping spherical coordinates $\bs{\theta} = (\theta_1,\dots, \theta_{p-1}) \in \mathbb{R}^{p-1}$ to Cartesian coordinates $\bs{\gamma} = (\gamma_1,\dots, \gamma_p) \in~\mathbb{R}^p$, such that 
  $\gamma_1 = \sin(\theta_1)$, $\gamma_l = \sin(\theta_l) \prod_{j=1}^{l-1} \cos(\theta_j),$ for $l \in  \{2,\dots, p-1\}$, and $\gamma_p = \prod_{j=1}^{p-1} \cos(\theta_j)$. We restrict each $\theta_j$ to the range $[-\pi/2, \pi/2]$, which ensures that the corresponding $\bs{\gamma} = q(\bs{\theta}) \in \mathbb{R}^p$ covers all $\bs{\gamma}$ satisfying $||\bs{\gamma}||_2 = 1$ and $\gamma_p \geq 0$, and avoids the inclusion of both $\bs{\gamma}$ and $-\bs{\gamma}$ in the parameter domain to prevent  posterior distribution multimodality. The sparsity of $\bs{\gamma}$ is directly related to the sparsity of $\bs{\theta}$ due to the correspondence between the two representations. When $\bs{\gamma}$ is sparse, the corresponding  $\bs{\theta}$ is also sparse, which allows us to promote the sparsity of $\bs{\gamma}$ using a sparsity-inducing prior on the  spherical coordinates~$\bs{\theta}$. 

\subsubsection{Prior on projection directions}
 We define a Spike-and-Slab Lasso (SSL) prior on $\bs{\theta}$, initially introduced by \citet{rovckova2018spike} in the context of  sparse mean estimation. The SSL prior is a mixture of two Laplace distributions with parameters $h_0$ and $h_1$, where $h_0 < h_1$. The ``spike" component has the smaller scale $h_0$, and the ``slab" component has the larger scale $h_1$.   The prior promotes sparsity via the ``spike" component, while the ``slab" component allows for nonsparse values in $\bs{\theta}$. It flexibly learns  the mixture proportions and the allocation of each variable ($\theta_j$) to either the ``spike" or  the ``slab" component. Assuming independent priors on the elements of $\bs{\theta}$,  for each $j \in  \{1,\dots, p-1\}$, the prior on $\theta_j$ is defined by 
\begin{align} \label{eq:prior:theta}
p(\theta_j|m_j) &=  \left\{\frac{1}{2h_0}\text{exp}\left(-\frac{|\theta_j|}{h_0}\right\}\right)^{m_j}  \left \{\frac{1}{2h_1}\text{exp}\left(-\frac{|\theta_j|}{h_1}\right)\right\}^{1-m_j}, \ 
p(m_j|w) = w^{m_j} (1-w)^{1-m_j}, 
w \sim Beta(\alpha_w, \beta_w).  
\end{align}
In \eqref{eq:prior:theta}, $p(\theta_j|m_j)$ defines a mixture of  Laplace distributions with parameters $h_0$ and $h_1$ respectively, $m_j$ is a binary variable indicating the allocation of each $\theta_j$ to either the ``spike" or the ``slab" component, and  $w \in (0, 1)$ is the mixing proportion.
The prior on  $\bs{\theta}$ defined through \eqref{eq:prior:theta} and the correspondence between $\bs{\theta}$ and $\bs{\gamma}$ implies the prior on $\bs{\gamma}$
\begin{equation} \label{eq:priorgamma}
p(\bs{\gamma}| \mb{m}) = |\text{det}(\mb{J}_{\theta})|^{-1} \prod_{j=1}^{p-1}p(\theta_j|m_j), \ \text{where} \  |\text{det} (\mb{J}_{\bs{\theta}})| =  \prod_{j=1}^{p-2} \cos(\theta_j)^{p-1-j},
\end{equation}
where $\mb{m} = (m_1,\dots, m_{p-1}) \in \mathbb{R}^{p-1}$, and $\mb{J}_{\theta}$ is the Jacobian matrix of the transformation from $\bs{\theta}$ to $\bs{\gamma}$.

\begin{remark} \label{remark:4}
The  prior \eqref{eq:priorgamma} does not explicitly encourage the last ($p$-th) element of $\bs{\gamma}$ (i.e. $\gamma_p$) to be exactly zero, which would require at least one $\theta_j$ to be $-\pi/2$ or $\pi/2$. However, it can still produce sparse estimates for $\gamma_p$ due to its multiplicative construction, since $\gamma_p$ is a product of $p-1$ cosine terms, and for large $p$ this product tends to shrink towards zero if some of the cosine terms are not exactly 1. Note that the ``slab" component of the SSL prior allows $\theta_j$ to take values far from 0, which can lead to a sparse estimate of $\gamma_p$ for large $p$. An alternative approach could use a mixture of three Laplace distributions, adding a third component centered at $-\pi/2$ or  $\pi/2$ to estimating a zero $\gamma_p$. However,  due to increased computational complexity and our focus on larger $p$, we choose not to pursue this option in this study. 
\end{remark}

 \subsubsection{Prior on ridge functions}
 To estimate the nonlinear function $g$ in \eqref{eq:priorgamma}, we use $J$ natural cubic B-spline basis functions $B_1, \dots , B_J$ and write $g$ as $g(u) = \sum_{j=1}^J c_j B_j(u)$. The interior knots of $B_j$'s are placed at the quantiles of the indices $\bs{\gamma}^T\mb{M}_i\bs{\gamma}$ for $i \in \{1,\dots, n\}$, with the boundary knots at  $\min_i(\bs{\gamma}^T\mb{M}_i\bs{\gamma})$ and $\max_i(\bs{\gamma}^T\mb{M}_i\bs{\gamma})$. Let  $\mb{B}_{j,\bs{\gamma}} = (B_j(\bs{\gamma}^T \mb{M}_1 \bs{\gamma}), \dots, B_j(\bs{\gamma}^T \mb{M}_n \bs{\gamma}))$ be the evaluations of the $j$-th basis function at the indices, and $\tilde{\bs{B}}_{\bs{\gamma}} = (\mb{B}_{1,\bs{\gamma}}, \dots, \mb{B}_{J,\bs{\gamma}}) \in \mathbb{R}^{n\times J}$ contain  evaluations from all basis functions. Following 
 \citet{antoniadis2004bayesian}, we specify a prior on the spline basis coefficients $\mb{c} = (c_1,\dots, c_J)$: 
 \begin{equation} \label{eq:prior:c}
 	\mb{c} \sim N \left( \mb{c}_{0, \bs{\gamma}}, \sigma^2 \bs{\Sigma}_{0, \bs{\gamma}}\right), \text{where} \  \mb{c}_{0, \bs{\gamma}} = \bs{\Sigma}_\rho \tilde{\bs{B}}_{\bs{\gamma}}^T \mb{r}, \  \bs{\Sigma}_\rho = \left(\tilde{\bs{B}}_{\bs{\gamma}}^T\tilde{\bs{B}}_{\bs{\gamma}} + \rho \mb{I}_J \right)^{-1}, \bs{\Sigma}_{0, \bs{\gamma}} = \left(\tilde{\bs{B}}_{\bs{\gamma}}^T\tilde{\bs{B}}_{\bs{\gamma}} \right)^{-1},
 \end{equation}
in which $\mb{r} = (r_1,\dots, r_n)$ is the vector of partial residual responses in \eqref{eq:model:sim},  $\mb{I}_J$ is the  $J \times J$ identity matrix, and  $\rho$ is a small positive constant that regularizes the spline fit.  For the variance parameter $\sigma^2$, we use the prior $\sigma^2 \sim IG(\alpha, \beta)$ as introduced in \Cref{subsec:backfitting}.

 \subsubsection{MCMC posterior sampling for sparse single-index model}
We derive the conditional posterior distributions and perform Gibbs sampling \citep{geman1984stochastic} to obtain posterior samples. For notational simplicity, in this section we do not explicitly write predictors $\mb{M}_i$'s in the conditioning part of the probabilities, as all distributions are conditioned on them. 

We can write the joint posterior distribution of model parameters based on the product of the likelihood and a prior \begin{align} \label{eq:joint}
 p(\bs{\gamma}, \mb{c}, \mb{m}, w, \sigma^2| \mb{r}) 
&\propto p(\mb{r}| \bs{\gamma}, \mb{c}, \sigma^2)p(\bs{\gamma}, \mb{c}, \mb{m}, w, \sigma^2) \nonumber \\ 
&\propto  p(\mb{r}| \bs{\gamma}, \mb{c}, \sigma^2 )p(\mb{c}|\bs{\gamma})p(\bs{\gamma}| \mb{m}) p(\mb{m}|w) p(w)p(\sigma^2).
\end{align}
and derive the conditional posterior distributions for $\mb{m}$, $w$, and $\bs{\gamma}$, respectively. 

First focusing on the factors that involve the ``spike" or 	``slab" allocation indicators $\mb{m}$ in \eqref{eq:joint},  we derive the conditional posterior for $\mb{m}$.  We let $\psi_{h}(u) = \text{exp}(-|u|/h)/(2h)$ denote the density of a Laplace distribution with parameter $h$ and mean 0, and derive the conditional posterior  

$$p(\mb{m}| \mb{r},  \bs{\gamma}, \mb{c},w, \sigma^2) \propto p(\bs{\gamma}| \mb{m})p(\mb{m}|w) 
\propto \prod_{j=1}^{p-1} \left ( \psi_{h_0}(\theta_j)^{m_j} \psi_{h_1}(\theta_j)^{1- m_j}  w^{m_j} (1-w)^{1-m_j} \right ), $$ which corresponds to the product of independent conditional distributions for each $m_j$:   
$m_j| \mb{r}, \bs{\gamma}, w \sim Bernoulli (\tilde{p}_j)$ with $\tilde{p}_j = w_j \psi_{h_0}(\theta_j)/\{w_j\psi_{h_0}(\theta_j) + (1-w_j)\psi_{h_1}(\theta_j)\}$.  Similarly, we derive the conditional posterior for the mixing proportion $w$ in \eqref{eq:joint} 
$$p(w| \mb{r},  \bs{\gamma}, \mb{c}, \mb{m}, \sigma^2) \propto p(\mb{m}|w)p(w) =  w^{\sum_{j=1}^{p-1} m_j} (1-w)^{\sum_{j=1}^{p-1} (1 - m_j)} w^{\alpha_w - 1} (1-w)^{\beta_w - 1},$$
which yields 
$ w| \mb{m} \sim Beta \left(\tilde{\alpha}_w , \tilde{\beta}_w  \right)$,  where $\tilde{\alpha}_w =  \sum_{j=1}^{p-1} m_j+ \alpha_w$, and $\tilde{\beta}_w = \sum_{j=1}^{p-1} (1 - m_j) + \beta_w.$  For the conditional posterior of the projection direction $\bs{\gamma}$, we follow the procedure in \citet{antoniadis2004bayesian} and marginalized over $\mb{c}$ and $\sigma^2$ to obtain  
\begin{align} \label{eq:posterior:gamma}
p(\bs{\gamma}| \mb{r}, \mb{m}, w)  &= \int_{\sigma^2} \left(\int_{\mb{c}}  p(\bs{\gamma}, \mb{c}, \sigma^2| \mb{m}, \mb{r}, w) d \mb{c} \right)   d\sigma^2 \nonumber \\ 
&\propto  p(\bs{\gamma}| \mb{m})   \int_{\sigma^2} \left(\int_{\mb{c}} p(\mb{r}| \bs{\gamma}, \mb{c}, \sigma^2 )p(\mb{c}|\bs{\gamma}) d \mb{c} \right) p(\sigma^2)  d\sigma^2 \nonumber 	 \\
&\propto p(\bs{\gamma}| \mb{m})  \left(S(\bs{\gamma}) + 2\beta \right)^{-\alpha - n/2},
\end{align}
where $p(\bs{\gamma}|\mb{m})$ is defined in \eqref{eq:priorgamma}, and $S(\bs{\gamma}) = \bs{r}^T \bs{r}  - \bs{r}^T\tilde{\mb{B}}_{\bs{\gamma}} \left( \bs{\Sigma}_{\rho} + \frac{1}{2}\bs{\Sigma}_{0, \bs{\gamma}}  - \frac{1}{2}\bs{\Sigma}_{\rho} \bs{\Sigma}_{0, \bs{\gamma}}^{-1}\bs{\Sigma}_{\rho} \right)  \tilde{\mb{B}}_{\bs{\gamma}}^T\bs{r}$.   The derivation details of \eqref{eq:posterior:gamma} are provided in Appendix B. 

To sample from  $p(\bs{\gamma}| \mb{r}, \mb{m}, w)$ in \eqref{eq:priorgamma}, 
we use the Metropolis-Hastings algorithm \citep{metropolis1953equation} with a von Mises-Fisher (vMF) proposal distribution for $\bs{\gamma}$. This vFM distribution is defined  with concentration parameter $\lambda$ and mean direction $\bs{\gamma}_{t-1}$ obtained from the  previous MCMC iteration in \Cref{alg:pbr}. Its density is proportional to $\text{exp}(\lambda \bs{\gamma}^T \bs{\gamma}_{t-1})$, where a larger $\lambda$ results in a distribution more concentrated around $\bs{\gamma}_{t-1}$.  We adapt the value of $\lambda$ based on the acceptance rate of the past iterations as introduced in \Cref{sec:simulation}. Let $\bs{\gamma}'$ be the proposed $\bs{\gamma}$. Due to the symmetry of the vMF distribution, i.e.  $p(\bs{\gamma}_{t-1}| \bs{\gamma}') = p(\bs{\gamma}'|\bs{\gamma}_{t-1})$, the acceptance probability in the Metropolis-Hastings algorithm can be simplified and is given by 
 $$p_{\text{acc}}(\bs{\gamma}', \bs{\gamma}_{t-1})=  \text{min}\left(1, \frac{p(\bs{\gamma}'| \mb{r}, \mb{m}_t, w_t)}{p(\bs{\gamma}_{t-1}| \mb{r}, \mb{m}_t, w_t)}\right),$$
where $p(\bs{\gamma}| \mb{r}, \mb{m}_t, w_t)$ is defined in \eqref{eq:posterior:gamma}. After obtaining a sample of $\bs{\gamma}_t$ through the Metropolis-Hastings algorithm, we follow \citet{antoniadis2004bayesian} by setting $\mb{c}_t = \mb{c}_{0, \bs{\gamma}_{t}}$ (see \eqref{eq:prior:c}), resulting in a sample of the ridge function $g$ in the single-index model \eqref{eq:model:sim}: $g_t(u) = \sum_{j=1}^J c_{j,t}B_j(u)$.  The proposed Metropolis-within-Gibbs sampling algorithm for generating posterior samples from the sparse single-index model is summarized in 
\Cref{alg:sim}. When applying \Cref{alg:sim}  at the $t$-th MCMC iteration in \Cref{alg:pbr} to sample the $k$-th additive component,  we take $\bs{\gamma}_{t-1}$ as $\bs{\gamma}_{k,t-1}$, and take $\bs{\theta}_{t-1}$ as the spherical coordinates of $\bs{\gamma}_{k,t-1}$ with the sign adjusted for its last element $\bs{\gamma}_{k,t-1, (p)}$, such that $q(\bs{\theta}_{t-1})= \text{sign}\left(\bs{\gamma}_{k,t-1, (p)}\right) \bs{\gamma}_{k,t-1}$.

 \SetKwFor{If}{If}{}{}
\SetKwFor{Else}{Else}{}{EndIf}

\begin{algorithm} \caption{Posterior sampling from Bayesian sparse single-index model} \label{alg:sim}

\Input{$\mathcal{R} = \{ (\mb{M}_i, r_i), i \in \{1,\dots, n\}\}$, $\lambda$, $\bs{\gamma}_{t-1}, \bs{\theta}_{t-1}, w_{t-1}$ }{}
\For{$j \in \{1, \dots, p-1\}$}{
Sample $m_{j,t} \sim  Bernoulli (\tilde{p}_j), \ \text{where} \  \tilde{p}_j = \frac{w_{t-1} \psi_{h_0}(\theta_{j,t-1})}{w_{t-1}\psi_{h_0}(\theta_{j, t-1}) + (1-w_{t-1})\psi_{h_1}(\theta_{j, t-1}) }$
}
Sample $w_t \sim Beta \left(\tilde{\alpha}_w , \tilde{\beta}_w  \right),  \ \text{where} \ \tilde{\alpha}_w =  \sum_{j=1}^{p-1} m_{j,t}+ \alpha_w, \ \tilde{\beta}_w = \sum_{j=1}^{p-1} (1 - m_{j, t}) + \beta_w $ \\
Sample $\bs{\gamma}' \sim $ vMF $(\lambda, \bs{\gamma}_{t-1})$\\
Set $p_{\text{acc}}(\bs{\gamma}', \bs{\gamma}_{t-1})=  \text{min}\left(1, \frac{p(\bs{\gamma}'| \mb{r}, \mb{m}_t, w_t)}{ p(\bs{\gamma}_{t-1}| \mb{r}, \mb{m}_t, w_t)}\right)$ \\
Sample $u \sim Unif(0, 1)$ \\
\If{$u < p_{\text{acc}}(\bs{\gamma}', \bs{\gamma}_{t-1})$}{$\bs{\gamma}_t = \bs{\gamma}'$} 
\Else{$\bs{\gamma}_t = \bs{\gamma}_{t-1}$}{}
\Return{$\mb{m}_t$, $w_t$, $\bs{\gamma}_t$}{}
\end{algorithm}

\section{Simulation} \label{sec:simulation}
We conducted extensive simulations to evaluate the performance of our  method, comparing it to alternative frequentist and Bayesian approaches. We assessed our method's predictive accuracy and parameter inference as a semi-parametric approach, and focused on prediction as a nonparametric approach.

\subsection{Setup} 

The symmetric matrix predictors for $N$ subjects were generated based on the eigenvalue decomposition  $
\mb{M}_i = \mb{A}_{i} \mb{E}_{i} \mb{A}_{i}^T$ for $i \in  \{1,\dots, N\},$ where  $\mb{A}_{i} \in \mathbb{R}^{p \times p}$ is the  matrix of eigenvectors and $\mb{E}_i= \text{Diag}(e_{i,1},\dots, e_{i,p})$ contains the corresponding eigenvalues.  We generated orthonormal $\mb{A}_{i}$'s independently  for each subject $i$ from a uniform distribution on the Stiefel manifold using the method of \citet{stewart1980efficient}  
implemented in 
 the \texttt{pracma} package \citep{pracma}. The eigenvalues  $e_{i,j}$'s were sampled
 i.i.d. from $\mathcal{U}(-10,10)$ for $i \in ~\{1,\dots, N\}$ and  $j \in \{1,\dots, p\}$.  

The responses ($Y_i$) were simulated 
based on two scenarios: the first  scenario  with a ``correctly specified" model~\eqref{eq:model} under different numbers ($K$) of additive terms and dimensions  ($p \times p$) of matrix predictors, and the second scenario  with a ``misspecified" model  where our method was used for prediction.  These scenarios are described below.   
\begin{itemize}
\item 	``Correctly specified" scenario: We generated $Y_i$ based on model \eqref{eq:model} with the variance $\sigma^2 = 1$. To simulate the projection directions $\bs{\gamma}_1, \dots, \bs{\gamma}_K$, we first sampled  an orthornormal matrix $\mb{V} \in \mathbb{R}^{p \times p}$ using the \texttt{pracma} package, and for  $k \in \{1,\dots, K\}$, we set $\tilde{\bs{\alpha}}_k =  \mb{V}\mb{q}_k$, where $\mb{q}_k = (q_{k,1},\dots, q_{k,p})^T$ with each element $q_{k,j}$ drawn i.i.d. from $\mathcal{U}(0,1)$. We then set $p-4$ randomly selected elements of $\tilde{\bs{\alpha}}_k$ to zero to obtain $\bs{\alpha}_k$, and let $\bs{\gamma}_k = \bs{\alpha}_k /||\bs{\alpha}_k||_2$.  We  considered four combinations of  $(p, K)$: $(15, 2)$, $(15, 3)$, $(15, 4)$, and $(25, 2)$. 
These combinations allow us to compare 
different complexities of the regression function and matrix sizes, evaluating how the method scales in terms of predictive performance and inference results. 
 The values $p = 15$ and $25$ match  the dimension of the matrix predictors available in the HCP dataset (see \Cref{sec:hcp}). We set $\mu = 0$, and the link functions $g_1(u) = -u + c_1$, 
 $g_2(u) = -u^2/4 + c_2$, $g_3(u) = 2 \text{exp}(-u/5) + c_3$, and $ g_4(u) = u^2/4 + c_4$, where the constants $c_1$ to $c_4$  were set to make $g_1$ to $g_4$ satisfy the identifiability conditions in \Cref{remark:3}. 
  \item ``Misspecified" scenario: We generated $Y_i$ based on $Y_i =  f_1(\langle \mb{M}_i, \mb{C} \rangle) +  f_2(\langle \mb{M}_i, \mb{C} \rangle) +  \epsilon_i$, where $f_1(u)= 2u$, $f_2(u)= 2u^2$, $\epsilon_i$ followed i.i.d. $N(0, \sigma^2)$ with $\sigma^2 = 1$, and the coefficient matrix $\mb{C} \in \mathbb{R}^{p \times p}$ was constructed as $\mb{C} = \mb{U} \mb{U}^T$ with $\mb{U} \in \mathbb{R}^{p \times r}$, where we first sampled an orthonormal $\tilde{\mb{U}} \in \mathbb{R}^{p \times r}$ using the \texttt{pracma} package  and then set randomly selected $p-8$ elements per its column to zero to form $\mb{U}$.  When $r = 1$, the model is ``correctly specified". We considered $r \in \{2, 3\}$ as ``misspecified" cases.  The value  of the dimension $p$ was set to $30$. \end{itemize}

For both scenarios, the training set consisted of $n = 400$ subjects and the test set consisted of 1000 subjects, totaling $N = 1400$. We repeated the experiment for 100 independent runs. In each run,  the projection directions $\bs{\gamma}_1$, \dots, $\bs{\gamma}_K$  for the ``correctly specified" scenario and the coefficient matrix $\mb{C}$ for the ``misspecified" scenario were generated independently.

\subsection{Implementation details}
We compared the following methods, including available alternatives (\textbf{PPR}, \textbf{BayesPPR}, \textbf{SAM}, \textbf{SBGAM}, \textbf{BTR}, and \textbf{BNTR}) and our proposed approach (\textbf{PBR(U)} and  \textbf{PBR(SS)}):

 \begin{itemize}
  \setlength\itemsep{0.1em}
\item \textbf{PPR}: projection pursuit regression \citep{friedman1981projection} 
\item \textbf{BayesPPR}: Bayesian projection pursuit regression \citep{collins2024bayesian} 
\item \textbf{SAM}: sparse additive models \citep{ravikumar2009sparse}
\item \textbf{SBGAM}: sparse Bayesian generalized additive models \citep{bai2022spike} 
\item  \textbf{BTR}: Bayesian tensor regression \citep{guhaniyogi2017bayesian}
\item \textbf{BNTR}: broadcasted nonparametric tensor regression \citep{zhou2024broadcasted}
\item\textbf{PBR(U)}: the proposed projection-pursuit Bayesian  regression with the uniform prior  
\item \textbf{PBR(SS)}: the proposed projection-pursuit Bayesian regression with the spike-and-slab prior 
\end{itemize}

  \textbf{PPR} and \textbf{BayesPPR} are projection pursuit regression methods for multivariate predictors:   \textbf{PPR} is the classical frequentist approach, while \textbf{BayesPPR}  is a  recent Bayesian method that incorporates a prior designed for variable selection. In both methods, the vectorized upper triangular portion of each $\mb{M}_i$, including diagonal elements, were used as features.  We fit \textbf{PPR} using the \texttt{stats} package \citep{Rsoftware} and selected the number of additive terms from $\{1,2,3,4\}$ via 5-fold cross-validation.  For \textbf{BayesPPR}, we used the implementation from the \texttt{BayesPPR} package \citep{BayesPPR} which automatically selected the number of additive terms.

 \textbf{SAM} and \textbf{SBGAM} are methods for  sparse additive models for multivariate predictors.   When applied to symmetric matrix predictors, both methods assume the model $E(Y_i|\mb{M}_i) = \sum_{j\geq k}f_{j,k}(\mb{M}_{i, (j,k)})$, where  $f_{j,k}: \mathbb{R} \rightarrow \mathbb{R}$ for $j \in \{1,\dots, p\}$,  $k \in \{j, \dots, p\}$ are  functions to be estimated.  \textbf{SAM} is a frequentist method that uses sparse backfitting to estimate $f_{j,k}$'s with univariate smoothers, while encouraging some $f_{i,k}$'s to be zero to promote sparsity.  
 We utilized the \texttt{SAM} package \citep{SAMpkg} with natural cubic splines to approximates each $f_{j,k}$, selecting the sparsity tuning parameter via 5-fold cross validation from 30  default values.    \textbf{SBGAM} is a Bayesian method that fits each $f_{i,k}$ with basis functions and applies a group sparsity-inducing prior (an extension of the SSL prior) on the vector of basis coefficients.   The implementation of \textbf{SBGAM} was adopted from the \texttt{sparseGAM} package \citep{sparseGAMpkg}, which includes a function  for performing cross-validation to select the hyperparameter that controls the sparsity of the fit.

  \textbf{BTR} is a Bayesian linear tensor regression method, assuming  $E(Y_i|\mb{M}_i) = \langle \mb{M}_i, \mb{C} \rangle $, using the CP decomposition \citep{kolda2009tensor} for the coefficient $\mb{C} \in \mathbb{R}^{p \times p}$. We implemented the model  from the author of \citet{guhaniyogi2017bayesian}'s Github repository \citep{btrgithub}, using 3000 warmup iterations and 3000 sampling iterations.  The rank of $\mb{C}$ was selected from $\{1,2,3,4\}$ based on the Watanabe-Akaike information criterion (WAIC) \citep{gelman2014understanding}.
  \textbf{BNTR} is a frequentist nonlinear tensor regression method that applies nonlinear functions ($f_1,\dots, f_R$) element-wise to the matrix predictor $\mb{M}_i$, generating features $\mb{F}_{i,r} \in \mathbb{R}^{p \times p}$ for $r \in \{1,\dots, R\}$.  The model assumes $E(Y_i|\mb{M}_i)  = \sum_{r = 1}^R \langle \mb{B}_r,  \mb{F}_{i,r} \rangle$, where $\mb{B}_r \in \mathbb{R}^{p \times p}$ is a rank-1 matrix. 
 We used the \texttt{BNTR} package \citep{BNTRpkg} with ridge penalization on $\mb{B}_r$'s  and estimated the $f_r$'s with a 5-knot truncated power basis. The value of $R$ was selected from $\{1,2,3,4\}$, and the regularization hyperparameter was chosen from 1 to 10 using 5-fold cross-validation.
  
  For all aforementioned alternative methods, the specified values or ranges for the parameters were chosen to ensure that further tuning would not significantly improve performance.  Parameters not explicitly mentioned were set to default values  in the respective implementations.

To illustrate the effects of the sparsity-inducing prior, we considered two variants of the proposed projection pursuit Bayesian regression (\textbf{PBR}): \textbf{PBR(U)} and \textbf{PBR(SS)}. \textbf{PBR(U)} uses a uniform prior $\mathcal{U}(-\pi/2, \pi/2)$ on each of $\bs{\theta}_{k,j}$ for $k \in \{1,\dots,K\}$ and $j \in \{1,\dots,p-1\}$ in the spherical coordinate representation of $\bs{\gamma}_k$'s. while  \textbf{PBR(SS)} uses the  spike-and-slab prior introduced in \Cref{subsec:sim}. We implemented the variants using the R software \citep{Rsoftware}, and specified a $N(0,3^2)$ prior on $\mu$ and an $IG(1,1)$ prior on $\sigma^2$, where $\mu$ and $\sigma^2$ are defined in model \eqref{eq:model}. The initial values of $\bs{\gamma}_k$'s in \Cref{alg:pbr} were based on a preliminary fit from \textbf{PPR}. Specifically, we vectorized the upper-diagonal portion (including diagonals) of each $\mb{M}_i$ and fit \textbf{PPR} with $K$ additive components using these vectorized predictors. The resulting coefficient vectors $\mb{a}_k \in \mathbb{R}^{p(p+1)/2}$ were then rearranged into a matrix $\mb{A}_k \in \mathbb{R}^{p \times p}$ to recover the coefficients for $\mb{M}_i$.  The $j$-th diagonal entry  $\mb{A}_{k, (j,j)}$ contains the element of $\mb{a}_k$ that corresponds to the coefficient for $\mb{M}_{i, (j,j)}$. The off-diagonal entries, $\mb{A}_{k, (j,j')} = \mb{A}_{k, (j',j)}$ for $j>j'$, correspond to the coefficient for $\mb{M}_{i, (j, j')}$ divided by 2.  The first eigenvector of $\mb{A}_k$ was then used as $\bs{\gamma}_{k,0}$.  Although other initial values can be used, our experiments show that this choice yielded good performance in terms of  convergence speed and quality of posterior samples. The concentration parameter $\lambda > 0$ in Algorithm \ref{alg:sim} was adaptively adjusted for each additive term based on the acceptance rate in the Metropolis-Hastings algorithm.  Initially set to 10000, $\lambda$ was multiplied by 1.1 if the acceptance rate in the previous 100 iterations was less than 20\% and divided by 1.1 if the rate exceeded 40\%. For the prior on ridge functions defined in \eqref{eq:prior:c}, we considered values of 0, 0.1, and 0.2 for $\rho$, and used the natural cubic spline basis  with $J \in \{2,4, 6\}$   functions  as $B_1,\dots, B_J$.



 For \textbf{PBR(SS)}, we initialized the mixing proportion $w$ at 0.5 in \eqref{eq:prior:theta}  and used a $Beta(1,1)$ prior for $w$.   Following the practice in \citet{bai2022spike},  we fixed  $h_1= 1$ and selected $h_0$ from a grid of values. For $p = 15$,  $h_0$ was chosen from $\{0.025,0.05,0.075,0.1\}$ and for $p = 25$,  $\{0.2, 0.3\}$ were added to the set for selection.  We selected the values of $\rho$ and $J$ for  \textbf{PBR(U)} and $\rho$, $J$, and $h_0$ for \textbf{PBR(SS)} to minimize the WAIC, and used 10000 warm-up iterations and 3000 sampling iterations for both \textbf{PBR} methods.

\subsection{Evaluation metrics} 

We evaluated the predictive performance in ``correctly specified" and ``misspecified" scenarios for all methods by calculating the mean-squared prediction error (MSPE) on the test data: $\text{MSPE} = \sum_{i \in \mathcal{I}}(\hat{Y}_i - Y_i)^2/|\mathcal{I}|$, where $\mathcal{I} = \{n+1, \dots, N\}$ denotes the subject indices for the test set, and $|\cdot|$ denotes the cardinality.  Among the Bayesian methods, \textbf{BayesPPR}, \textbf{BTR}, and \textbf{PBR} computed $\hat{Y}_i$ as the posterior mean of the predicted response for the $i$-th subject, and \textbf{SBGAM} computed $\hat{Y}_i$ based on the maximum a posteriori (MAP) estimation  suggested in \citet{bai2022spike}.  For the frequentist methods (\textbf{PPR}, \textbf{SAM}, and \textbf{BNTR}), $\hat{Y}_i$ was computed using point estimates of the model parameters.

When assessing the inference performance of \textbf{PBR}, in the ``correctly specified" scenario, we aligned the order and sign of the post-warmup samples ($\bs{\gamma}_{k,t}$'s for MCMC iteration $t \in \mathcal{T} = \{T', \dots , T\}$) to account for model identifiability. Specifically, in each of the 100 runs, we reordered $\bs{\gamma}_{k,t}$'s based on their absolute cosine similarity (ACS) with the $\bs{\gamma}_k$'s used for data generation.  For any $k, k' \in \{1, \dots , K\}$, the ACS between $\bs{\gamma}_{k',t}$ and $\bs{\gamma}_{k}$ is defined as $\text{ACS}(\bs{\gamma}_{k',t}, \bs{\gamma}_{k}) = |\bs{\gamma}_{k',t}^T\bs{\gamma}_{k}|$.  For $k \in \{1, \dots, K\}$, we computed the index $b_k = \argmax_{k' \in \mathcal{I}/\mathcal{I}_{k-1}} \sum_{t \in \mathcal{T}} \text{ACS}( \bs{\gamma}_{k',t},\bs{\gamma}_k)$, where $\mathcal{I} = \{1, \dots, K\}$,  $\mathcal{I}_k = \mathcal{I}_{k-1} \cup \{b_k\}$, and $\mathcal{I}_0 = \{\ \}$, such that $\bs{\gamma}_{b_k, t}$ was considered a posterior sample of $\bs{\gamma}_k$ at $t$.    To  adjust the sign of each $\bs{\gamma}_{b_k,t}$, we let $l^{\ast} = \text{arg max}_l(|\bs{\gamma}_{k,(l)}|)$ where $\bs{\gamma}_{k,(l)}$ is the $l$-th element of $\bs{\gamma}_{k}$. For  $t \in \mathcal{T}$, we compared the signs of the $l^{\ast}$-th element of $\bs{\gamma}_{b_k,t}$ (denoted as  $\bs{\gamma}_{b_k,t, (l^{\ast})}$ ) and $\bs{\gamma}_{k, (l^{\ast})}$, and if they disagreed, we multiplied $\bs{\gamma}_{b_k,t}$ by -1.

We then evaluated the quality of the posterior samples of $\bs{\gamma}_k$'s by computing $\text{ACS}(\bs{\gamma}_{b_k, t} , \bs{\gamma}_k)$ for $k \in \{1, \dots, K\}$ and $t \in \mathcal{T}$.  To further assess the element-wise coverage of $\bs{\gamma}_k$ using the posterior samples, we constructed 80\%  credible intervals for $\bs{\gamma}_{k,(l)}$,  denoted  as $\left [\bs{\gamma}_{k,(l)}^L, \bs{\gamma}_{k,(l)}^U \right]$,  where ``L"
and ``U" represent the lower and upper bounds, and defined the corresponding coverage of $\bs{\gamma}_{k,(l)}$ computed with the credible intervals as $\text{CoverCI}(\bs{\gamma}_{k,(l)}) = I \left( \bs{\gamma}_{k,(l)} \in \left[\bs{\gamma}_{k,(l)}^L, \bs{\gamma}_{k,(l)}^U \right] \right)$. Note that wider intervals are more likely to cover the true values, and thus $\text{CoverCI}$ alone does not fully reflect the precision of the estimates. Therefore, we also checked the length of credible intervals, denoted as $\text{LenCI}(\bs{\gamma}_{k,(l)}) =  \bs{\gamma}_{k,(l)}^U  - \bs{\gamma}_{k,(l)}^L$. 



\subsection{Results} 

\begin{figure} \centering
\includegraphics[scale=0.8]{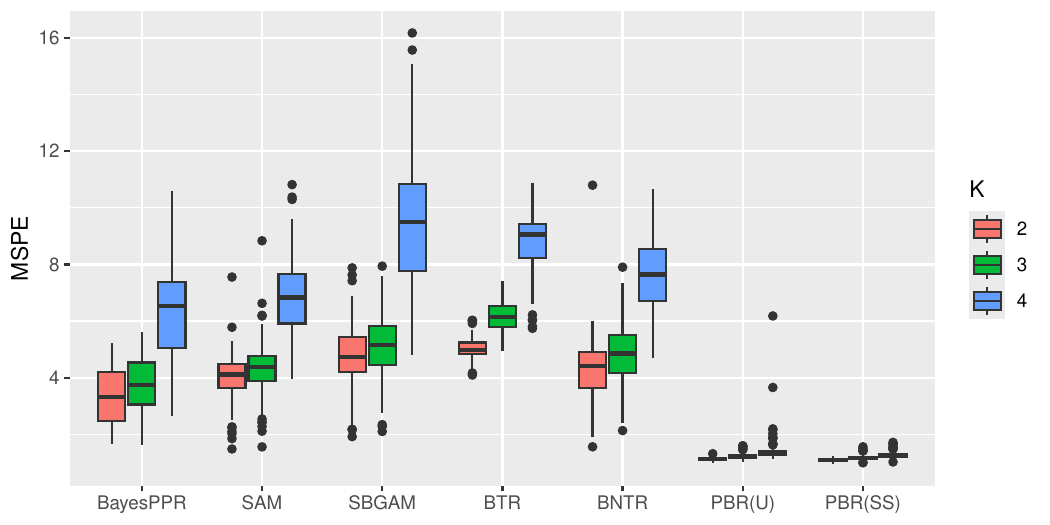}

\caption{Mean-squared prediction error (MSPE) on test sets for settings with $p = 15$ and $K \in \{2, 3, 4\}$ from 100 runs of the experiment in the ``correctly specified" scenario. }
\label{fig:mspe.K}
\end{figure}

Fig. \ref{fig:mspe.K} shows the MSPE on the test sets for each method, obtained from 100 runs of the experiment for $p = 15$ and across $K = 2, 3$ and 4.  Results from
\textbf{PPR} are not included due to its poor performance in all settings, producing the highest testing errors.  In Fig. \ref{fig:mspe.K}, the proposed \textbf{PBR(SS)} method outperforms all others, showing the lowest MSPE and smallest variance across all settings. As $K$ increases, the predictive performance worsens for all methods,  while \textbf{PBR(SS)} showing the least degradation. 
 Upon further examination of the training errors (see the Supplementary Materials \footnote{https://github.com/xmengju/PBR/Supplementary\_Material}),  \textbf{BTR} and \textbf{BNTR} yielded the highest training errors, despite incorporating the matrix structure in their tensor regression models.  This is due to the limited flexibility of their models, either assuming a linear model or applying the same nonlinear transformations to all matrix elements. Comparing the two \textbf{PBR} methods,  \textbf{PBR(SS)} produced slightly lower prediction errors compared to \textbf{PBR(U)}, although this difference is not visually apparent in the figure due to the scale of the y-axis.  When $K = 4$, \textbf{PBR(U)} generated some outliers of large MSPEs.

\begin{figure} \centering
\includegraphics[scale=0.7]{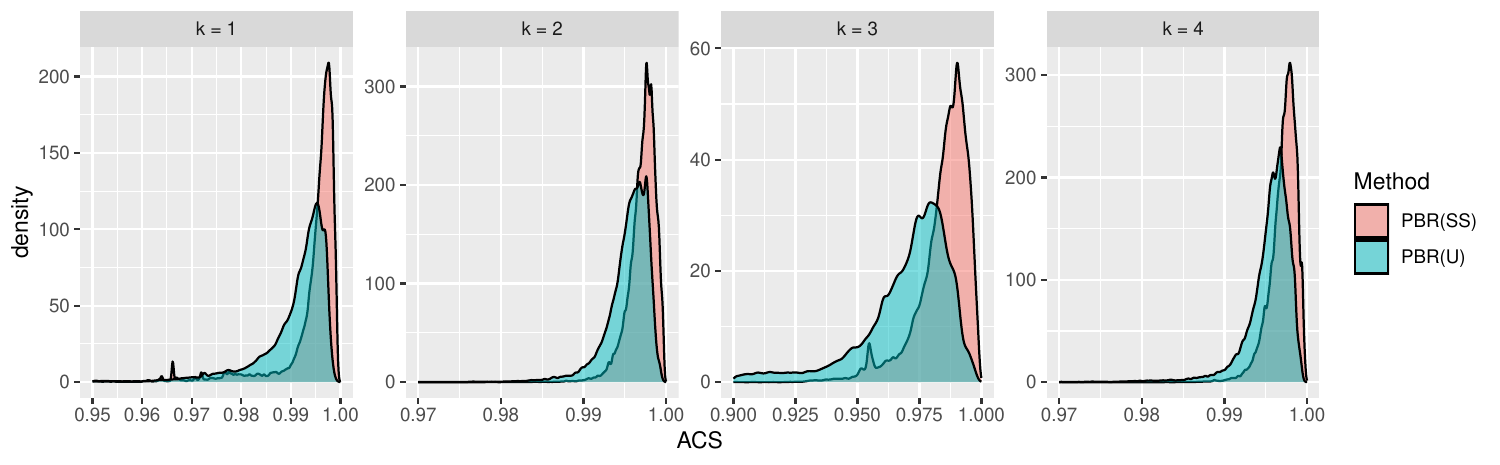}
\caption{Density of the absolute cosine similarity (ACS) of $\bs{\gamma}_k$'s for the $p = 15$ and $K = 4$ setting from 100 runs of the experiment. }
\label{fig:acs}
\end{figure}

Fig.  \ref{fig:acs} presents the case of $p = 15$ and $d = 4$, the most complex model amongst all,  to examine the inference results. The results for the other settings are provided in the Supplementary Materials, and the conclusions drawn from this setting are consistent with those from the other settings.  We pooled together the aligned samples  for $\bs{\gamma}_k$'s across the 100 runs and plotted the density of $\text{ACS}(\bs{\gamma}_{b_k, t} , \bs{\gamma}_k)$ for each $k \in \{1,\dots, K\}$ in Fig. \ref{fig:acs}.   The x-axis in each panel of  Fig. \ref{fig:acs} is left truncated to better visualize the differences in the bulk of the densities.   Compared to \textbf{PBR(U)}, \textbf{PBR(SS)} produced higher ACS values with more concentrated distributions,  suggesting that these samples more accurately reflect the true projection directions.  The performance of \textbf{PBR(SS)} is stable across different $k$ values, whereas the performance of \textbf{PBR(U)} varies, and notably worse for the exponential term ($k = 3$)  that contains a weaker signal (i.e. lower variance of the corresponding additive component across training subjects). 

\begin{figure} \centering
\includegraphics[scale=0.5]{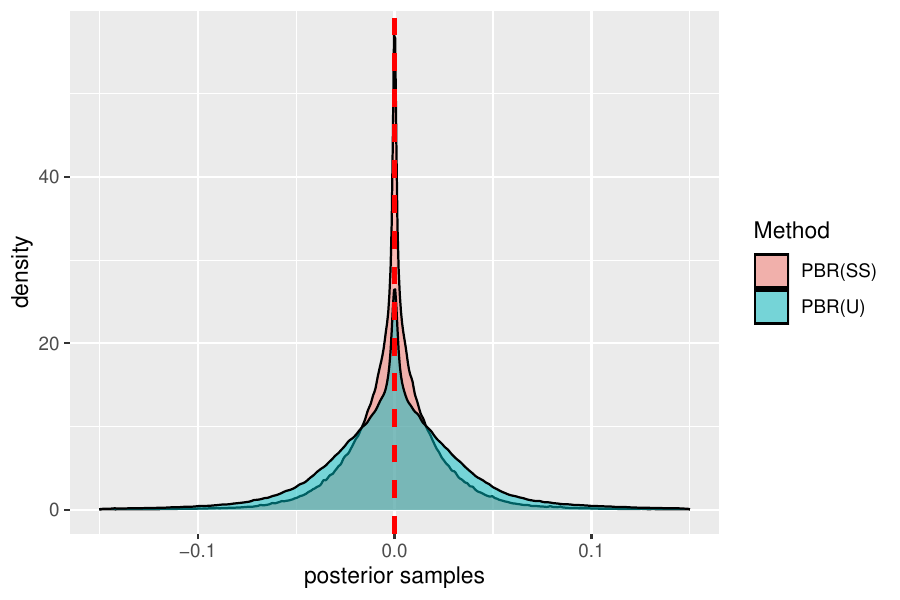}
\caption{Density of posterior samples for the true zero elements of $\bs{\gamma}_k$'s for the $p = 15$, $K = 4$ setting, generated with samples aggregated for $k\in \{1,\dots, K\}$ from 100 runs of the experiment. Vertical dashed line indicates a value at zero. }
\label{fig:density}
\end{figure}

To evaluate the effectiveness of the SSL prior in producing sparse estimates,  we compared the  distribution of the posterior samples corresponding to the true 
 zero entries in $\bs{\gamma}_k$'s. These samples were aggregated across different $k$'s over 100 runs. As illustrated in Fig. \ref{fig:density}, \textbf{PBR(SS)} produced sparser estimates compared to \textbf{PBR(U)}, characterized by thinner tails and a shaper spike at 0.

\begin{figure}
\centering
\hspace{0.1cm}
\begin{subfigure}[b]{0.49\textwidth}
   \includegraphics[width=1.05\linewidth]{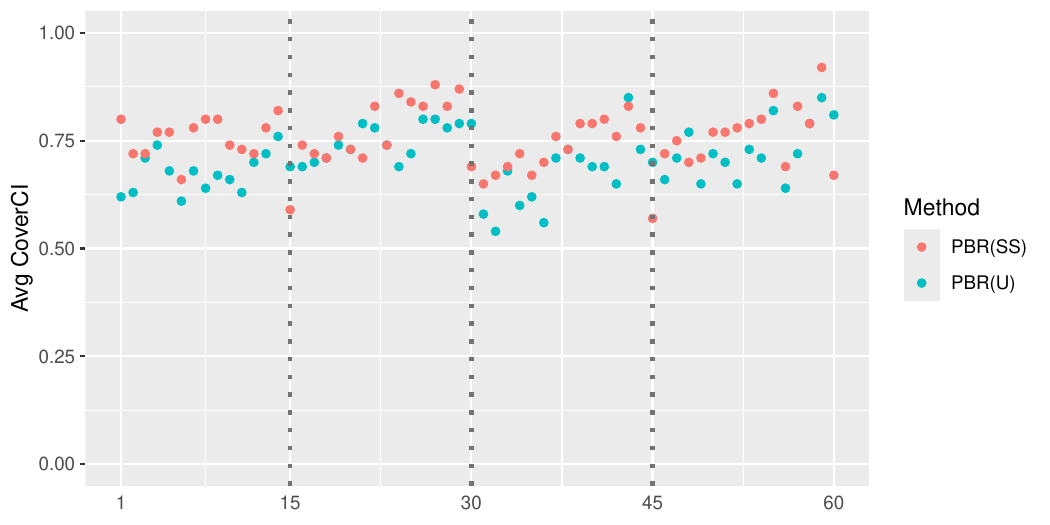}
   \caption{Average CoverCI}
\end{subfigure}
\begin{subfigure}[b]{0.49\textwidth}
\hspace{0.2cm}
   \includegraphics[width=1.05\linewidth]{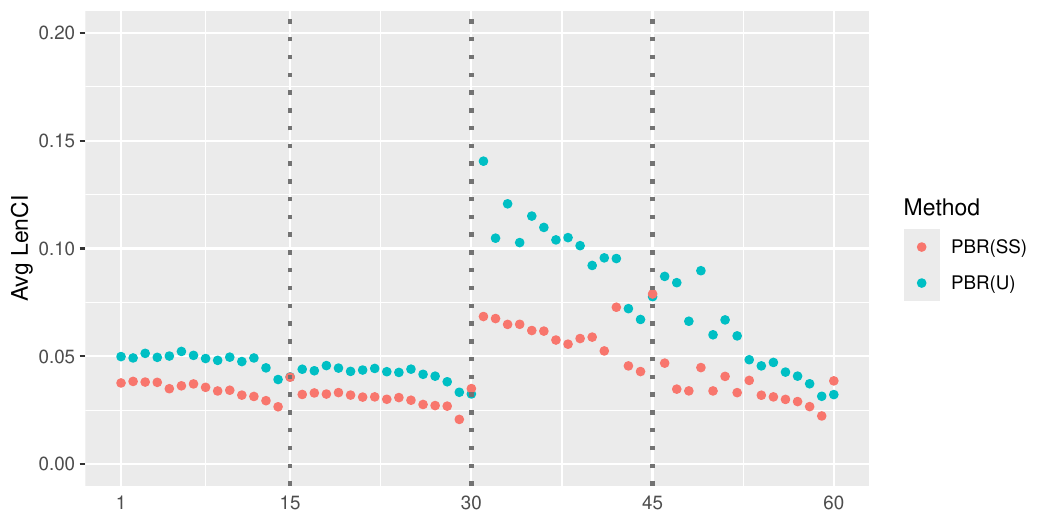}
   \caption{Average LenCI}
\end{subfigure}
\caption{CoverCI and LenCI for the $p = 15$ and $K = 4$ setting for each element of $\bs{\gamma}_k$'s, averaged from 100 runs of the experiment. }
\label{fig:coverCI}
\end{figure}

 For each entry $\bs{\gamma}_{k,(l)}$, we computed the coverage and length of 80\% credible intervals (CoverCI and LenCI, respectively) and averaged the values across 100 runs for each metric.   For easier comparison under each metric, we concatenated these values across  $\bs{\gamma}_{k,(l)}$'s, resulting in a vector of length $p \times K = 60$ for  \textbf{PBR(U)} and  \textbf{PBR(SS)} respectively, plotted in Fig. \ref{fig:coverCI}. Dotted vertical lines in the figure divide the vector for each $k \in \{1,2,3,4\}$. Panel~(a) in Fig.~\ref{fig:coverCI} shows that \textbf{PBR(SS)}, compared to \textbf{PBR(U)}, produced higher average CoverCI for most entries of $\bs{\gamma}_k$'s.  Although \textbf{PBR(SS)} resulted in lower coverage of the last element of $\bs{\gamma}_{k}$ (i.e. $\gamma_{k, (p)}$)  per $k$ compared to the other elements for reasons addressed in \Cref{remark:4}, the coverage rate for these elements are still higher compared to the lowest coverages produced by \textbf{PBR(U)}.  Meanwhile,  panel (b) in Fig. \ref{fig:coverCI}  suggests that  the LenCI generated by \textbf{PBR(SS)} are generally narrower compared to \textbf{PBR(U)}.  It is notable that the lengths are wider for $k = 3$ for both methods, as the corresponding exponential term has the weakest signal among the other additive terms.

\begin{figure}
\centering
\begin{subfigure}[b]{1\textwidth}
   \includegraphics[height=5cm, width = 16cm]{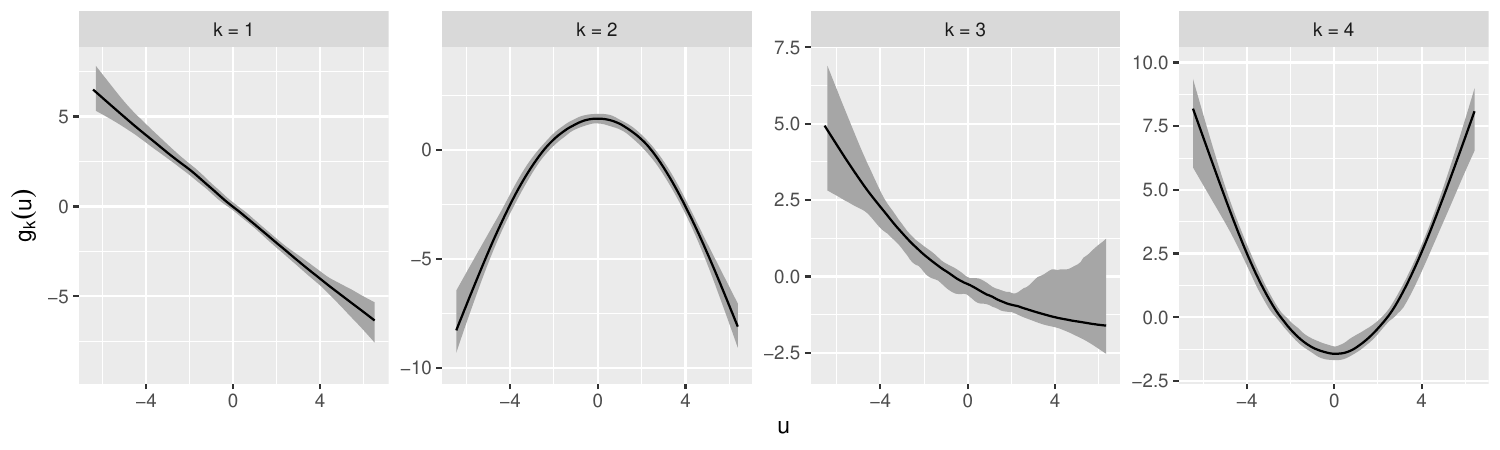}   \caption{PBR(U)}
\end{subfigure}

\begin{subfigure}[b]{1\textwidth}
   \includegraphics[height=5cm, width = 16cm]{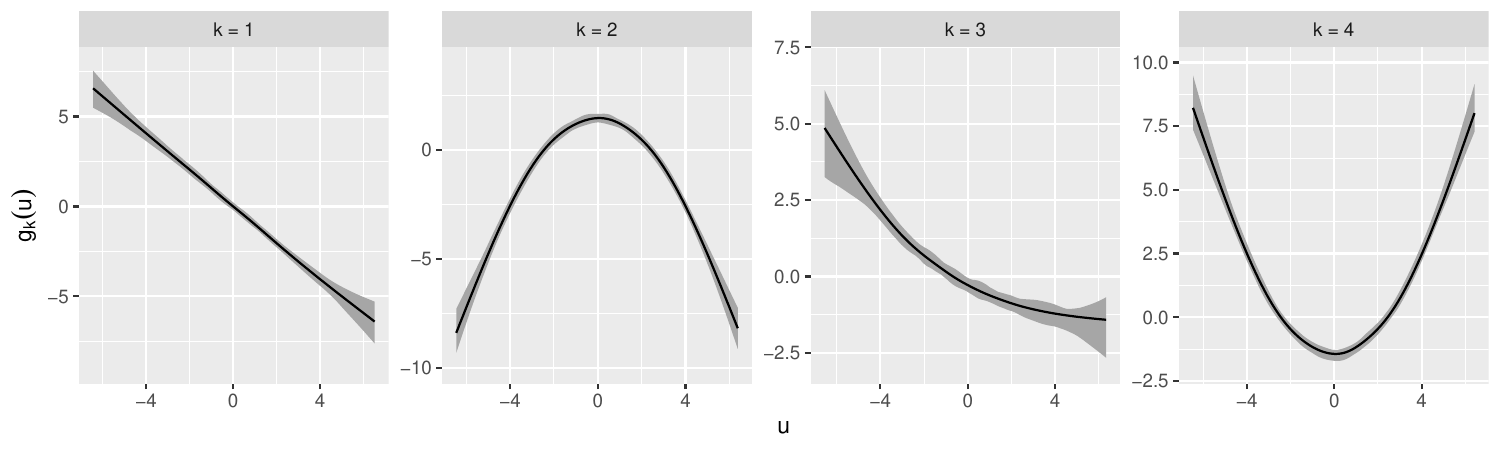}    \caption{PBR(SS)}
\end{subfigure}
\caption{Median and 80\% credible intervals for the posterior samples of ridge functions ($g_k$'s) generated by \textbf{PBR(U)}  and \textbf{PBR(SS)} for the $p = 15$ and $K = 4$ setting, from 100 runs of the experiment in the ``correctly specified" scenario. }
 \label{fig:ridge} 
\end{figure}

To compare the posterior samples of the ridge function $g_k$, we evaluated them at a common grid of indices. In Fig.~\ref{fig:ridge}, at each grid point, we plotted the median (solid line) and 80\% credible intervals (shaded areas) of the posterior samples $g_{k,t}$. The common grid was constructed using 100 evenly spaced data points within the range computed from the indices constructed with the true $\bs{\gamma}_k$, i.e. $\min_i \langle  \mb{M}_i, \bs{\gamma}_k\bs{\gamma}_k^T\rangle$ to $\max_i \langle  \mb{M}_i, \bs{\gamma}_k\bs{\gamma}_k^T\rangle$. 
Comparing the results from \textbf{PBR(U)} and \textbf{PBR(SS)} in Fig. \ref{fig:ridge}, both methods produced medians that accurately recover the true ridge function, but \textbf{PBR(SS)} yielded narrower credible intervals.

\begin{figure}
\centering
	   \includegraphics[scale=0.67]{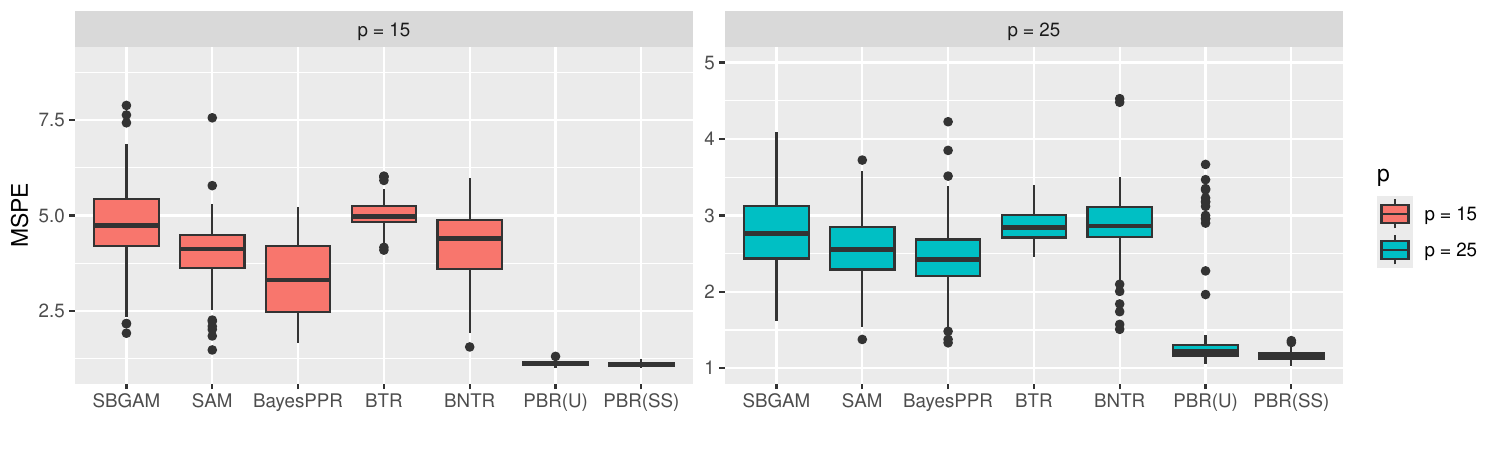}
	  \caption{Mean-squared prediction error (MSPE) on test sets for settings with $K=2$ and $p  \in \{15, 25\}$, from 100 runs of the experiment in the ``correctly specified" scenario. }
\label{fig:different:p}
\end{figure}

With $K$ fixed at 2, we further compared the predictive performance of different methods in Fig. \ref{fig:different:p} for two settings: $p = 15$ and $p = 25$. As $p$ increases, the performance of \textbf{PPR} (not shown) deteriorates notably due to overfitting. The other competitors performed better since the range of the index is smaller in our setup with $p = 25$, resulting in components that are easier to estimate. In  Fig. \ref{fig:different:p}, \textbf{PBR(SS)} shows consistently better performance compared to \textbf{PBR(U)},  which produced some outliers as $p$ increases. In the $p = 25$ setting, the results regarding the coverage, length of credible intervals, and estimation of ridge functions were similar to those observed in the $p = 15$ setting (see the Supplementary Materials). \textbf{PBR(SS)} consistently produced better inference results, in terms of higher coverage rate and narrower credible intervals for $\bs{\gamma}_k$ and $g_k$, compared to \textbf{PBR(U)}.

\begin{figure}
\centering
	 	 \includegraphics[scale=0.67]{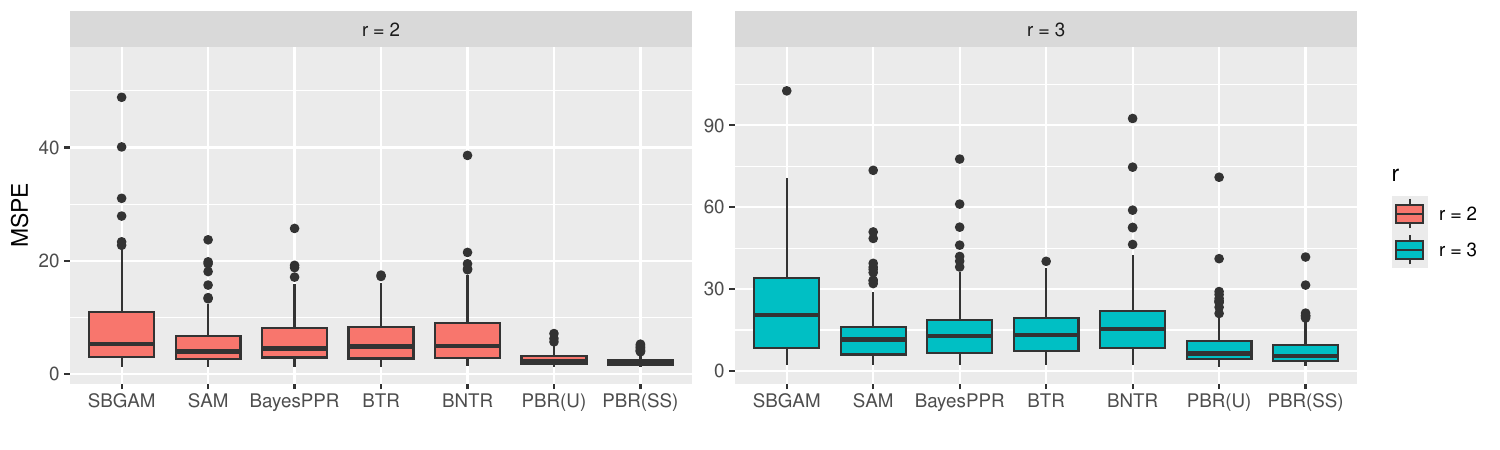}
	 \caption{Mean-squared prediction error (MSPE) on test sets for settings with $p = 30$ and $r \in \{2, 3\}$, from 100 runs of the experiment in the ``misspecified" scenario.}
	\label{fig:npsetting}
\end{figure}

Finally, we compared different methods in the ``misspecified" case, where our method was used as a nonparametric estimator for prediction. Among the tested $K$ values, $K = 2$ and 3 yielded the best performance for $r = 2$ and 3 settings respectively, and  thus the corresponding results are reported here.  In Fig. \ref{fig:npsetting}, for both settings, \textbf{PBR(SS)} shows the lowest test MSPEs with the smallest standard deviations, and \textbf{PBR(U)} performed the second best,   demonstrating the flexibility of the \textbf{PBR} proposal in scenarios that deviate from the assumed model.

\section{Application} \label{sec:hcp}
In this section we present an application of the proposed \textbf{PBR} method to data from the Human Connectome Project (HCP), using  brain region connectivity represented as SPD matrices to predict cognitive scores. Our analysis focuses on the resting-state functional magnetic resonance imaging (rs-fMRI) data from the S1200 release \citep{van2013wu}, which include behavioural and imaging data collected from healthy young adults. For each subject, the rs-fMRI data were collected across four  15-minute scanning sessions, with 1200 timepoints per session.  The rs-fMRI data was preprocessed according to \citet{smith2013functional}.

 We converted the rs-fMRI data into time series mapped onto spatial brain regions using a data-driven parcellation based on group spatial independent component analysis (ICA) \citep{mckeown1998analysis,calhoun2001method} with $p = 15$ and $25$ brain regions (i.e. ``network nodes")  described  as ICA maps \citep{filippini2009distinct},  denoted  as Nodes 1 to $p$.  The output,  available in the PTN (Parcellation + Timeseries + Netmats) format \footnote{https://www.humanconnectome.org/study/hcp-young-adult}, includes data for $N=1003$ subjects from the S1200 release \cite{van2013wu}.   Considering the temporal dependency in the time series , we performed thinning of the observed data based on the effective sample size (ESS) for each subject $i$, following the procedure described in \citet{park2024bayesian}:
$\text{ESS}(i) = \min_{ j \in \{1, \dots, p\}} T_i/ \left(1+ 2\sum_{t=1}^{T_i - 1} \text{cor} \left(x_{i,j}(t), x_{i,j}(1 +~ t)\right)\right),$
where  $x_{i,j}(t)$ denotes the signal at time $t$ in region $j$ for subject $i$.  The  $\text{ESS}(i)$ values range from 181 to 763, with an average of 349 across subjects.  We performed thinning  by subsampling the $\text{EES}(i)$ time points for each subject $i$ which were then used to  compute the covariance matrices $\mb{M}_i$'s.

\begin{figure}
\centering
	 \includegraphics[scale=0.4]{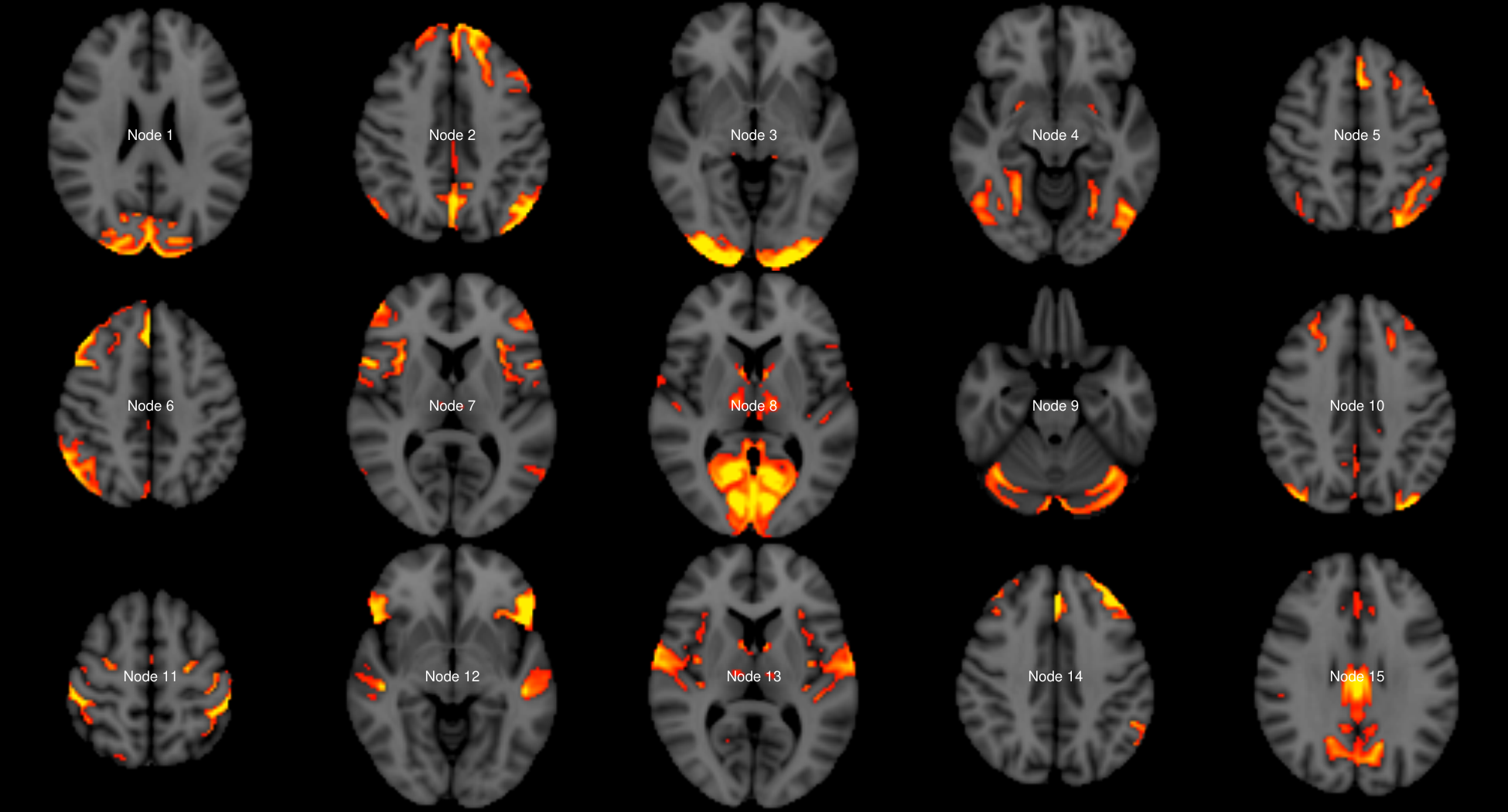}
	 \caption{The 15 independent components (ICs) from spatial group-ICA. The components are denoted as Node 1 to 15 (corresponding to 15 IC ``network nodes"), provided by the PTN dataset form the S1200 HCP release, represented at the most relevant axial slices in the MNI152 space. According to \citet{seiler2017multivariate}, these IC networks correspond to default network (Node 15), cerebellum (Node 9), visual areas (Node 1, Node 3, Node 4, and Node 8), cognition-language (Node 2, Node 5, Node 10, and Node 4), perception-somesthesis-pain (Node 2, Node 6, Node 10, and Node 14), sensorimotor (Node 7 and Node 11), executive control (Node 12) and auditory functions (Node 12 and Node 13). }
	\label{fig:brainimage}
\end{figure}

The response variable is the age-adjusted early childhood composite score (CogEarlyComp\_AgeAdj), which summarizes measures of cognitive functions evaluated by the NIH Toolbox for Assessment of Neurological and Behavioral Function  \citep{gershon2013nih,weintraub2013cognition}.  This composite score is calculated by averaging the normalized scores from four tests: Picture Vocabulary, Flanker,  Dimensional Change Card Sort, and Picture Sequence Memory, which assess cognitive functions of language comprehension,  inhibitory control, executive function, and episodic memory, respectively.   For detailed information on each test, see \citet{weintraub2013cognition}. To help specify sensible priors for the \textbf{PBR} methods, 
we standardized the response variable to have mean zero and unit variance. We used  a $N(0,1)$ prior for $\mu$ in model \eqref{eq:model}, and the other priors and implementation details remained consistent with those described in \Cref{sec:simulation}.

We divided the data into a training set of size 800, and used the remainder as test data.   The experiment was repeated for 50 runs, each with a different random data split.   For the competitor approaches, the implementation details were the same as described in \Cref{sec:simulation}.  For  \textbf{PBR} methods, we considered $\rho \in \{0, 0.1, 0.2\}$, $J \in  \{3, 4, 5\}$ (corresponding to 1,2, and 3 interior knots), and for  \textbf{PBR(SS)}  $h_0 \in \{0.1, 0.2,0.3,0.4\}$. To  optimize these parameters, we further divided the training data into a validation training set of size 700 and a validation test set of size 100. We selected the parameters ($\rho$, $h_0$, and $J$) that minimized the average MSPE on the validation test sets across all runs when trained with the validation training sets.   The final model was then fitted using the 
800 training observations and evaluated on the remaining test data. 
 Below, we report results for $p = 15$ where the corresponding brain regions have well-established functional interpretations in the literature \citep{seiler2017multivariate}. The most relevant axial slices for each region in the MNI-152 space \citep{evans19933d,mazziotta1995probabilistic,mazziotta2001probabilistic} are displayed in Fig.~\ref{fig:brainimage} along with their functions as introduced in \citet{seiler2017multivariate}. We tested $K = 2$ and $K = 3$ additive terms and found no significant difference in the predictive performance for \textbf{PBR} methods. Therefore, we report results for the more parsimonious model with $K = 2$.   Results for $p = 25$ are provided in the Supplementary Materials.

\begin{figure}
\centering
	 \includegraphics[scale=0.75]{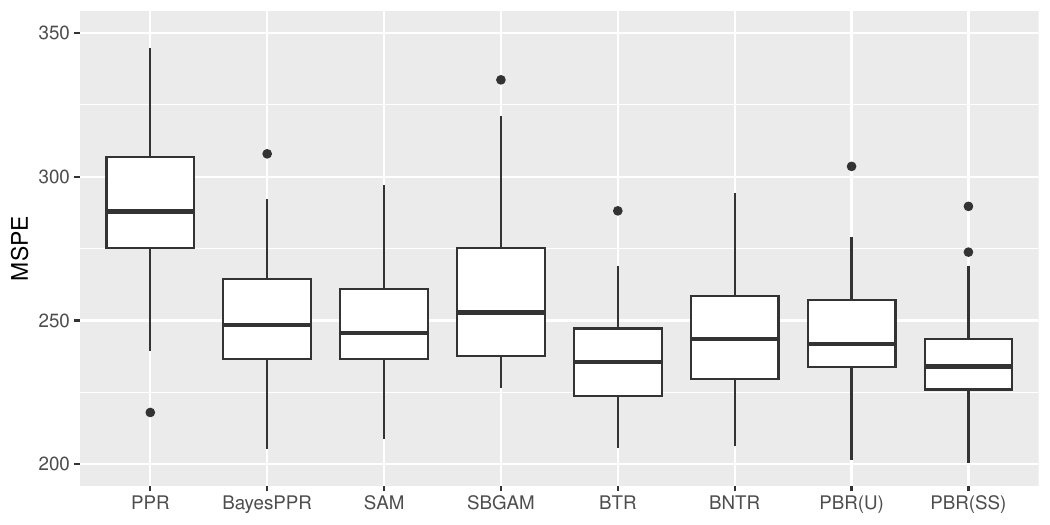}
	 \caption{Mean-squared prediction error (MSPE) on test sets obtained from 50 runs, each with a random data split. }
	\label{fig:hcpmspe}
\end{figure}

Fig. \ref{fig:hcpmspe}  shows prediction performance comparison of different methods in terms of
the MSPEs on test sets from 50 random data splits  for $p = 15$, computed with  responses of the original scale, recovered from standardization. \textbf{PBR(SS)} and \textbf{BTR} are the top performers, with  \textbf{PBR(SS)} exhibiting a smaller variance and a marginally better median of the MSPEs. 
 Generally,  tensor (or matrix)-based methods (\textbf{BTR}, \textbf{BNTR}, and \textbf{PBR}) outperform the other methods (\textbf{PPR}, \textbf{SBGAM}, \textbf{SAM} and \textbf{BayesPPR}) designed for vector inputs. Among all methods, \textbf{PPR} performs the worst, likely due to its inability to incorporate matrix structure and insufficient regularization.  The superior performance of \textbf{PBR(SS)} over \textbf{PBR(U)} highlights the benefits of accounting for the sparsity in projection directions ($\bs{\gamma}_k$'s).


The posterior samples of the ridge functions were largely monotone, with one increasing and one decreasing.  The model of \textbf{PBR} is identifiable up to the additive components' orders and the signs of the projection directions (see \Cref{theorem:1}).  To align the order of 
projection directions ($\bs{\gamma}_k$'s) and ridge functions ($g_k$'s) for each run, we  computed the  indices $u_{i,t} = \langle \mb{M}_i, \bs{\gamma}_{k,t}\bs{\gamma}_{k,t}^T \rangle $ and evaluated their correlation with $g_{k,t}(u_{i,t})$ for training subjects.  If this correlation was negative, we designated $g_{k,t}$ and $\bs{\gamma}_{k,t}$ as samples of  $g_1$ and  $\bs{\gamma}_1$, respectively. Otherwise, we assigned them as samples of $g_2$ and  $\bs{\gamma}_2$.

\begin{figure}
\centering
	 \includegraphics[scale=0.8]{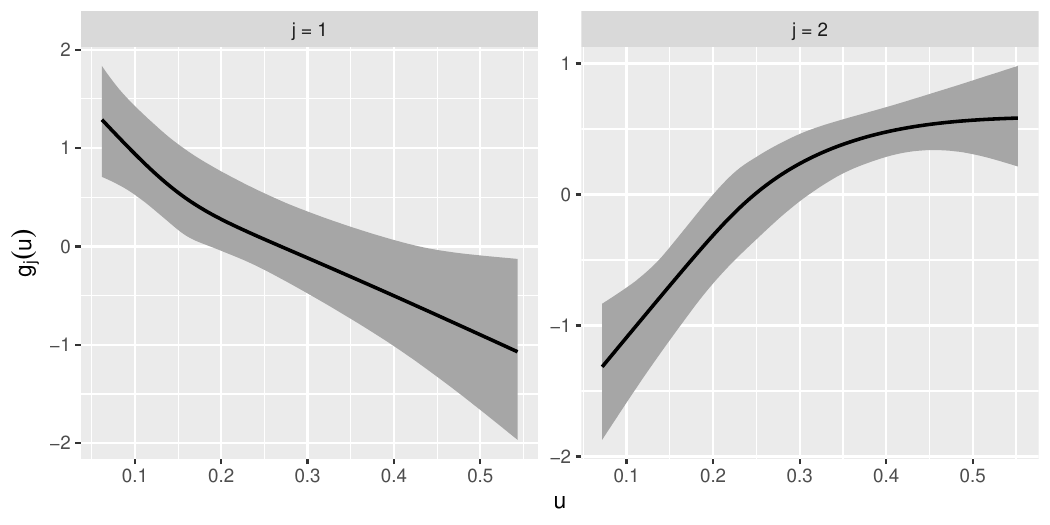}
	 \caption{Median and 80\% credible intervals for the posterior samples of ridge functions ($g_1$ and $g_2$) generated by \textbf{PBR(SS)} aggregated across 50 runs, each with a random data split.}
	\label{fig:hcpridge}
\end{figure}

Fig.~\ref{fig:hcpridge} presents the median (solid lines) and 80\% credible intervals (shaded areas) of the ridge functions evaluated on a common grid with the aligned posterior samples. The grid was constructed using evenly spaced points within the range defined by first  calculating the lower bound, $\min_{i, t} \langle  \mb{M}_i, \bs{\gamma}_{k,t}\bs{\gamma}_{k,t}^T\rangle$, and upper bound, $\max_{i,t} \langle  \mb{M}_i, \bs{\gamma}_{k,t}\bs{\gamma}_{k,t}^T\rangle$, for each run, then taking the median of these bounds across 50 runs.  
The median ridge functions in Fig.~\ref{fig:hcpridge} are generally monotonic and can be reasonably approximated by linear functions, explaining the good performance of the linear method \textbf{BTR}.   Additionally, $g_1$ shows  a larger variation across indices and wider credible intervals compared to $g_2$.

\begin{figure}
\centering
\hspace{0.1cm}
\begin{subfigure}[b]{0.49\textwidth}
\hspace{-0.5cm}
   \includegraphics[width=1.15\linewidth,page=1]{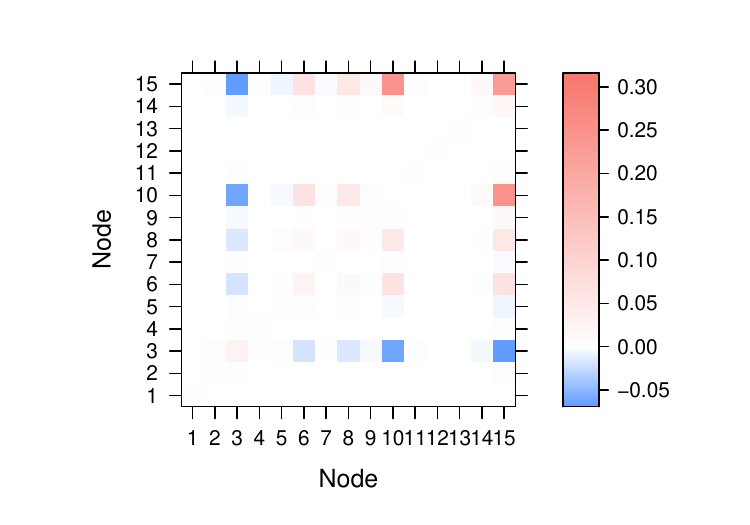}
   \caption{$\bs{\gamma}_1\bs{\gamma}_1^T$}
\end{subfigure}
\begin{subfigure}[b]{0.49\textwidth}
\hspace{-0.6cm}
   \includegraphics[width=1.15\linewidth,page=2]{Figures/CogEarlyComp_AgeAdj_hcp_gammas_white_s.pdf}
   \caption{$\bs{\gamma}_2\bs{\gamma}_2^T$}
\end{subfigure}
\caption{Median of the posterior samples of $\bs{\gamma}_1\bs{\gamma}_1^T$ and $\bs{\gamma}_2\bs{\gamma}_2^T$ for $p = 15$ generated by \textbf{PBR(SS)} aggregated across 50 runs, each with a random data split.}
\label{fig:hcp:gammas}
\end{figure}

In Fig.~\ref{fig:hcp:gammas}, we plotted the median of the rank-1  matrix coefficients ($\bs{\gamma}_k\bs{\gamma}_k^T$'s) constructed with the projection directions.  The 80\% credible intervals of these matrix coefficients are included in the Supplementary Materials. The entries with large-scale coefficients in Fig.~\ref{fig:hcp:gammas} primarily involve connectivities between Nodes 1, 3, 4, 5, 10, and 15.  According to \citet{seiler2017multivariate}, Nodes 1, 3, and 4 are associated with visual areas, Nodes 5 and 10 with cognitive-language functions, and Node 15 with the default network.  This finding aligns with our goal of predicting cognitive scores as the cognitive tests include assessments of language abilities and visual tasks.  Notably, the largest coefficient corresponds to the sample variance of the fMRI signals at Node 5, suggesting that the activeness of cognitive-language functions  plays an important role in predicting cognitive scores.




\section{Discussion} \label{sec:discussion}

In this paper, we propose a Bayesian projection-pursuit method for regression with symmetric matrix predictors. 
Our approach builds an additive multi-index estimator and servers as a supervised dimension reduction tool for symmetric matrix-valued data. Unlike existing methods that rely on restrictive assumptions on the matrix predictors or the regression functions, our method is highly flexible and yields interpretable projection directions  with uncertainty quantifications.  The sparsity-inducing prior for the projection directions effectively avoids overfitting and  reflects real-world scenarios where only a subset of matrix entries is predictive of the response.

The simulation study demonstrated that the proposed \textbf{PBR(SS)} method achieves superior predictive performance and inference results.  In analyzing the HCP data, \textbf{PBR(SS)} not only provides excellent  prediction accuracy  but also identifies key brain connections for predicting cognition scores. The improved results of \textbf{PBR(SS)} over its non-sparse counterpart (\textbf{PBR(U)})  show the effectiveness of sparse sampling for the projection directions.  Our method can also include additional real-valued  predictors  as additive terms with nonlinear ridge functions. This extension has been implemented in our  Github repository \footnote{https://github.com/xmengju/PBR}.

 Our proposal opens several directions for further work. We chose the number of additive terms~($K$) and the hyperparameter ($h_0$) in the sparsity-inducing prior for \textbf{PBR} using information criteria (e.g. WAIC) or validation procedures. Further research could  explore automatic selection methods based on the training data, such as applying reversible jump MCMC (RJMCMC) for $K$ as suggested by \citet{collins2024bayesian} or incorporating a hyperprior on $h_0$. Extending our approach to asymmetric matrices and higher-order tensor predictors is also of interest.
   

\section*{Acknowledgments}
This work was supported by the National Institute of Health (NIH Grant No.5 R01 MH099003). Data were provided by the Human Connectome Project, WU-Minn Consortium (Principal Investigators: David Van Essen and Kamil Ugurbil; 1U54MH091657) funded by the 16 NIH Institutes and Centers that support the NIH Blueprint for Neuroscience
Research; and by the McDonnell Center for Systems Neuroscience at Washington University.

\section*{Appendices}

\subsection*{Appendix A: Proof of \Cref{theorem:1}}

\setcounter{equation}{0}
\renewcommand{\theequation}{A.\arabic{equation}}

Below we provide the proof for \Cref{theorem:1} in \Cref{subsection:identifiability}.  

\begin{lemma} \label{eq:lemma1}
(\citet{khatri1968solutions}) 
 Consider the functional equation 
$$\phi_1(\mb{c}_1^T\mb{t}) + \cdots + \phi_r(\mb{c}_r^T\mb{t})= \xi_1(t_1) + \cdots + \xi_p(t_p)$$
defined for $|t_i| \leq R,$ for some $R \in \mathbb{R}^{+}$ and $i \in \{1,\dots, p\}$, where $\mb{t} = (t_1,\dots, t_p)^T$ and $\mb{c}_1, \dots , \mb{c}_r$ are column vectors of a $p \times r$ matrix $\mb{C}$ with $p \geq r$. If $\mb{C}$ has full column rank and each column has at least two non-zero entries, then $\phi_1, \dots, \phi_r$ and $\xi_1, \dots , \xi_p$ are all quadratic functions. 
\end{lemma}

\Cref{eq:lemma2} below states Cauchy's additive functional equation \cite{aczel1966lectures} which is used in the proof. 
\noindent 
\begin{lemma} \label{eq:lemma2}
If 
$$g(x) + g(y) = g(x+y)$$
holds for $x, y \in \mathbb{R}$ and a continuous function $g$, then $g$ is linear. 
\end{lemma}

\noindent 
\textbf{Proof of Theorem 1}:  
Setting $\mb{M} = \mb{0}_{p \times p}$,  it follows from \eqref{eq:identify} that $\mu = \tilde{\mu}$, proving (a) in \Cref{theorem:1}. In the following,  the proof is divided into Part (1) and Part (2). In Part (1), we prove \Cref{theorem:1} under \Cref{assumption:1} and  Condition $(\ast)$ defined below, using techniques from \citet{yuan2011identifiability}.  \\

\noindent 
\textbf{Condition $(\ast)$}:  Not all functions in $\{g_1, \dots, g_K\}$  and $\{\tilde{g}_1, \dots,  \tilde{g}_{\tilde{K}}\}$  are quadratic. \\

  In Part (2), we prove \Cref{theorem:1}  when Condition $(\ast)$ is violated.  \\

\noindent 
\textbf{Part (1)}: Given that $\bs{\Gamma} = (\bs{\gamma}_1, \dots, \bs{\gamma}_K) \in \mathbb{R}^{p \times K}$ is column full-rank, as  per  \Cref{assumption:1} (c), we construct a matrix $\mb{F} = (\bs{\Gamma}, \bs{\Gamma}^{\perp}) \in \mathbb{R}^{p \times p}$, s.t.  $\bs{F}^T\mb{F} = \mb{I}_p$. Let $\mb{Q} = \mb{F}^{-1}\tilde{\bs{\Gamma}} \in \mathbb{R}^{p \times \tilde{K}}$ for
 $\tilde{\bs{\Gamma}} = (\tilde{\bs{\gamma}}_1, \dots , \tilde{\bs{\gamma}}_{\tilde{K}}) \in \mathbb{R}^{p \times \tilde{K}}$, with its columns $\mb{q}_j =  (q_{j,1}, \dots, q_{j,p})^T = \mb{F}^{-1} \tilde{\bs{\gamma}}_j$, and $||\mb{q}_j||_2= 1$ by construction, for $j \in \{1, \dots , \tilde{K}\}$. Under this notation, $\tilde{\bs{\gamma}}_j^T\mb{M}\tilde{\bs{\gamma}}_j= \mb{q}_j^T \mb{F}^T\mb{M} \mb{F}\mb{q}_j = \mb{q}_j^T\mb{Z}\mb{q}_j$ where  $\mb{Z} = \mb{F}^{T}\mb{M} \mb{F} \in \mathbb{R}^{p \times p}$.  Denoting the diagonal elements of $\mb{Z}$ as $z_1, \dots, z_p$, we rewrite \eqref{eq:identify} as  
 \begin{equation} \label{eq:identify:appendix}
  g_1(z_1) + \cdots + g_K(z_K) = \tilde{g}_1(\mb{q}_1^T\mb{Z} \mb{q}_1) +  \cdots + \tilde{g}_{\tilde{K}}(\mb{q}_{\tilde{K}}^T\mb{Z} \mb{q}_{\tilde{K}}), 
\end{equation}
\noindent 
which holds for any $\mb{Z}$ that can be generated from $\mb{M} \in \text{Sym}^p$. In particular, \eqref{eq:identify:appendix} holds for any diagonal $\mb{Z}$ since the corresponding $\mb{M} = (\mb{F}^{-1})^T \mb{Z} \mb{F}^{-1} \in \text{Sym}^p$.

The rest of the proof proceeds by induction. First, consider the case of $\tilde{K} = 1$. We will show that if
\begin{equation} \label{eq:0}
  g_1(z_1) +  \cdots + g_K(z_K) =  \tilde{g}_1(\mb{q}_1^T\mb{Z} \mb{q}_1),
\end{equation}
then $K = 1$, $g_1 = \tilde{g}_1$, and $\bs{\gamma}_1 = (-1)^l \tilde{\bs{\gamma}}_1$ for $l \in \{0, 1\}$.  To prove this, for any 
$j \in \{1, ..., p\}$, let $\mb{Z}$ be a diagonal matrix with $z_{j} > 0$, and $z_k = 0$ for $k \neq j$.
If \eqref{eq:0} holds, then 
\begin{equation} \label{eq:1}
	\tilde{g}_1(q_{1,j}^2z_j) = 
	\begin{cases}
		g_j(z_j) & 	 j \leq K \\
		0 & K< j \leq p.
	\end{cases}
\end{equation}
It follows from \eqref{eq:1} that $q_{1,j} = 0$ for $K < j \leq p$ due to a contradiction: 
if there exist a $j$ such that $K < j \leq p$, and  $q_{1,j} \neq 0$,   \eqref{eq:1} would imply $\tilde{g}_1 \equiv 0$, contradicting \Cref{assumption:1} (a). Furthermore, since $||\mb{q}_{1}||_2  = 1$, there is at least one  $j \leq K$ such that $q_{1,j} \neq 0$ . Below we establish that there can only be one $j \leq K$ such that $q_{1,j} \neq 0$.

By contradiction,  assume there exist indices $j' \leq K$ and $j'' \leq K$, such that   $q_{1,j'} \neq 0$ and  $q_{1,j''} \neq 0$. Let $z_j' >0$, $z_j'' >0$, and $z_k = 0$ for $k \notin \{j', j^{''}\}$. From \eqref{eq:0} and \eqref{eq:1}, we have
\begin{equation} \label{eq:3}
  g_1( z_{j'}) + g_2(z_{j''}) = \tilde{g}_1(q_{1,j'}^2 z_{j'} + q_{1,j''}^2 z_{j''}) = \tilde{g}_1( q_{1,j'}^2 z_{j'}) +   \tilde{g}_1( q_{1,j''}^2 z_{j''}).
\end{equation}
 Applying \Cref{eq:lemma2} to \eqref{eq:3},  $\tilde{g}_1$ is a linear function, and $g_1$ and $g_2$ are also linear, contradicting \Cref{assumption:1}~(b). Thus, $\mb{q}_{1}$ only has one nonzero element (assumed to be the $j'$-th element). Since  $||\mb{q}_1||_2 = 1$, we have $|q_{1,j'}| = 1$, and  $\bs{F}\mb{q}_{1} = (-1)^l\bs{\gamma}_{j'}$ for $l \in  \{0, 1\}$.  By the definition  $\mb{q}_1 = \mb{F}^{-1}\tilde{\bs{\gamma}}_1$, it follows that $\mb{F}\mb{q}_1 = \tilde{\bs{\gamma}}_1 = (-1)^l\bs{\gamma}_{j'}$. Based on  $|q_{1,j'}| = 1$, from \eqref{eq:1},  we find $\tilde{g}_1(z_{j'}) = g_{j'}( z_{j'})$ and $0 = \tilde{g}_1(q_{1,j}^2 z_{j}) = g_{j}(z_{j})$ for $j \neq j'$. Without loss of generality, we let $j' = 1$ and have proved identifiability for $\tilde{K} = 1$.

To invoke induction, we assume that if 
\begin{equation}\label{eq:4.0}
g_1(z_1) + \cdots + g_K(z_K) =  \tilde{g}_1(\mb{q}_1^T\mb{Z} \mb{q}_1) +  \cdots + \tilde{g}_{\tilde{K}}( \mb{q}_{\tilde{K}}^T\mb{Z} \mb{q}_{\tilde{K}}), 
\end{equation}
where $\mb{q}_1, \dots, \mb{q}_{\tilde{K}}$ are linearly independent, then $K = \tilde{K}$, and there exists a permutation $\pi(1), \dots, \pi(K) $  of $\{1, \dots, K\}$, such that $g_j = \tilde{g}_{\pi(j)}$, and $\mb{q}_{\pi(j)}/q_{\pi(j), j}$ is the $j$-th column of the $p \times p$ identity matrix $\mb{I}_p$, denoted as $\mb{e}_j$, for $j \in \{1, \dots, K\}$. This implies that there exist $l_j \in \{0, 1\}$   such that 
$\bs{\gamma}_j = (-1)^{l_j} \tilde{\bs{\gamma}}_{\pi(j)}$ for $j \in \{1, \dots, K\}$.

For  $\tilde{K} >1$   consider 
\begin{equation}  \label{eq:4}
g_1(z_1) + \cdots  + g_K(z_K) =  \tilde{g}_1(\mb{q}_1^T\mb{Z} \mb{q}_1) +  \cdots + \tilde{g}_{\tilde{K}}( \mb{q}_{\tilde{K}}^T\mb{Z} \mb{q}_{\tilde{K}})  + \tilde{g}_{\tilde{K}+1}( \mb{q}_{\tilde{K}+1}^T\mb{Z} \mb{q}_{\tilde{K}+1}),
\end{equation}
where, without loss of generality, $K \geq \tilde{K} + 1$.  Apply \Cref{eq:lemma1} by setting $\mb{t} = (z_1, \dots, z_p)$, $\xi_j = g_j$ for $j \in  \{1, \dots, K\}$, $\xi_{K+1}, \dots, \xi_{p} \equiv 0$, $r = \tilde{K}+1$, and $\phi_j = \tilde{g}_j$, $\mb{c}_j = (q_{j,1}^2, \dots, q_{j,p}^2)$ for $j\in \{1, \dots, \tilde{K}+1\}$.  
From \Cref{assumption:1} (c) that $\tilde{\bs{\Gamma}}$ is column full-rank, it can be inferred that $\mb{c}_1, \dots,  \mb{c}_{\tilde{K}+1}$ are linearly independent.   \Cref{eq:lemma1} implies that  Condition ($\ast$) is violated if each column of $\bs{C}$ has at least two non-zero entries.  Consequently,  there must exist a $j' \in \{1, \dots, \tilde{K}+1\}$, such that $\mb{q}_{j'}$  has only one non-zero entry (it cannot have all zero entries since $||\mb{q}_{j'}||_2 = 1$). By permutation of the order of the coordinates and $\mb{q}_j$'s (and the corresponding $\tilde{g}_j$'s and $\tilde{\bs{\gamma}}_j$'s), we let $j' = \tilde{K} + 1$ and  $q_{\tilde{K}+1, K}$ be the only non-zero entry of $\mb{q}_{\tilde{K}+1}$. 

For $j \in  \{1, \dots, \tilde{K}\}$, define $\bs{\eta}_j = \mb{q}_j - q_{j,K} \mb{q}_{\tilde{K}+1}.$ By construction, $\bs{\eta}_1, \dots, \bs{\eta}_{\tilde{K}}$ are linearly independent since $\mb{q}_1, \dots, \mb{q}_{\tilde{K}+1}$ are linearly independent. Setting $z_{K} = 0$,  from \eqref{eq:4},  we have \begin{equation*}
  g_1(z_1) + \cdots  + g_{K-1}(z_{K-1}) =  
 \tilde{g}_1(\bs{\eta}_1^T\mb{Z} \bs{\eta}_1) +  \cdots + \tilde{g}_{\tilde{K}}(\bs{\eta}_{\tilde{K}}^T\mb{Z} \bs{\eta}_{\tilde{K}}). 
\end{equation*}
 By induction, from \eqref{eq:4.0} we conclude that $K - 1= \tilde{K}$, and for $j \in \{1, ..., K-1\}$,  $g_j = \tilde{g}_{\pi(j)}$, and $\bs{\eta}_{\pi(j)}/\eta_{\pi(j), j} = \mb{e}_j$.  Without loss of generality, we let $\pi(j) = j$ and rewrite \eqref{eq:4} as 
\begin{align} \label{eq:5}
 g_1(z_1)  + \cdots + g_K(z_K)   &=  \tilde{g}_1( q_{1,1}^2z_1 +  q_{1,K}^2z_K) +  \cdots  + \tilde{g}_{\tilde{K}}(q_{\tilde{K},\tilde{K}}^2z_{\tilde{K}} + q_{\tilde{K},K}^2z_K) +  \tilde{g}_K(z_K).
\end{align}
 Let $z_j = 0$ for $j \notin \{1, K\}$ in \eqref{eq:5}, we obtain 
\begin{align} \label{eq:6}
g_1(z_1)  +  g_K(z_K)  = \tilde{g}_1( q_{1,1}^2z_1 +  q_{1,K}^2z_K) + \tilde{g}_2( q_{2,K}^2z_K) + \cdots  + \tilde{g}_{\tilde{K}}( q_{\tilde{K},K}^2z_K) +  \tilde{g}_K(z_K). 
\end{align}
Setting $z_j = 0$ for $j \neq 1$ in \eqref{eq:5} yields
\begin{equation} \label{eq:appendix:1}
	g_1(z_1)   = \tilde{g}_1( q_{1,1}^2z_1 ). 
\end{equation}
Setting $z_j = 0$ for $j \neq K$ in \eqref{eq:5} gives 
\begin{equation} \label{eq:7}
g_K(z_K)   = \tilde{g}_1( q_{1,K}^2z_K) + \tilde{g}_2( q_{2,K}^2z_K) +  \cdots  + \tilde{g}_{\tilde{K}}( q_{\tilde{K},K}^2z_K) +  \tilde{g}_K(z_K).
\end{equation}
Substituting $g_1(z_1)$ from \eqref{eq:appendix:1} and $g_K(z_K)$ from \eqref{eq:7} into  \eqref{eq:6}, we obtain 
$$\tilde{g}_1( q_{1,1}^2z_1 + q_{1,K}^2z_K) = \tilde{g}_1(q_{1,1}^2z_1) + \tilde{g}_1(q_{1,K}^2z_K).$$ If $q_{1,K} \neq 0$, \Cref{eq:lemma2} implies that $g_1$ and $\tilde{g}_1$ are linear functions. Using the same technique, we can show that for $j \in \{2, \dots, K-1\}$ if  $q_{j,K} \neq 0$, \Cref{eq:lemma2} implies that $g_{j}$ and $\tilde{g}_{j}$ are linear functions. Based on \Cref{assumption:1} (b), there cannot be more than one linear function in $g_1, \dots , g_K$. Hence, among $q_{1,K}, \dots, q_{K-1,K}$, there is at most one element that is nonzero. 
 If there exists  $j' \in \{1, \dots,  K-1 \}$ such that $q_{j',K} \neq 0$, it follows  from $ \tilde{\bs{\gamma}}_{j'}=\mb{F} \mb{q}_{j'}$  that  $\tilde{\bs{\gamma}}_{j'}$ is a linear combination of $\bs{\gamma}_{j'}$ and $\bs{\gamma}_{K}$, which would contradict the full rank condition of $\tilde{\bs{\Gamma}}$  in \Cref{assumption:1} (c). Therefore $\mb{q}_j/q_{j,j} = \mb{e}_j$, implying  $\bs{\gamma}_{j} = (-1)^{l_{j}} \tilde{\bs{\gamma}}_{j}$ for $l_j  \in \{0, 1\}$ and $j \in \{1, \dots K\}$.   Lastly, for any $j \in \{1, \dots, K\}$, by setting $z_j = 0$ for all $j' \neq j$ in \eqref{eq:5}, we conclude that $g_j = \tilde{g}_j$.  \\

\noindent 
\textbf{Part (2)}: We now prove \Cref{theorem:1} under the violation of Condition ($\ast$).  Let $g_k(u) = a_{k, 1}u^2 + a_{k, 2}u$ for $k \in \{1, \dots, K\}$, and $\tilde{g}_{l}(u) = b_{l, 1}u^2 + b_{l, 2}u$ for $l \in \{1, \dots, \tilde{K}\}$ and write \eqref{eq:identify}  as 
\begin{align} \label{eq:part2:1}
\sum_{k = 1}^K a_{k,1} (\bs{\gamma}_k^T\mb{M}\bs{\gamma}_k)^2 + \sum_{k = 1}^K a_{k,2} (\bs{\gamma}_k^T\mb{M}\bs{\gamma}_k) = \sum_{l = 1}^{\tilde{K}} b_{j,1} (\tilde{\bs{\gamma}}_l^T\mb{M}\tilde{\bs{\gamma}}_l)^2 + \sum_{l = 1}^{\tilde{K}} b_{j,2} (\tilde{\bs{\gamma}}_l^T\mb{M}\tilde{\bs{\gamma}}_l).
\end{align}
Without loss of generality, assume $K \geq \tilde{K}$. Taking the derivative of \eqref{eq:part2:1} with respect to $\mb{M}$ yields
\begin{align} \label{eq:part2:2}
2 \sum_{k = 1}^K a_{k,1} (\bs{\gamma}_k^T\mb{M}\bs{\gamma}_k)\bs{\gamma}_k\bs{\gamma}_k^T + \sum_{k = 1}^K a_{k,2} (\bs{\gamma}_k\bs{\gamma}_k^T) = 2 \sum_{l = 1}^{\tilde{K}} b_{l,1} (\tilde{\bs{\gamma}}_l^T\mb{M}\tilde{\bs{\gamma}}_l)\tilde{\bs{\gamma}}_l\tilde{\bs{\gamma}}_l^T + \sum_{l = 1}^{\tilde{K}}  b_{l,2} (\tilde{\bs{\gamma}}_l\tilde{\bs{\gamma}}_l^T). 
\end{align}
Setting $\mb{M} = \mb{I}_p$ in \eqref{eq:part2:2} gives
\begin{equation} \label{eq:part2:3}
\sum_{k=1}^K (2a_{k,1} + a_{k,2}) \bs{\gamma}_k \bs{\gamma}_k^T =  \sum_{l=1}^{\tilde{K}} (2b_{l,1} + b_{l,2}) \tilde{\bs{\gamma}}_l \tilde{\bs{\gamma}}_l^T. 
\end{equation}
Setting $\mb{M} = \mb{0}_p$  in \eqref{eq:part2:2} gives
\begin{equation} \label{eq:part2:4}	
\sum_{k=1}^K  a_{k,2} \bs{\gamma}_k \bs{\gamma}_k^T =  \sum_{l=1}^{\tilde{K}}  b_{l,2} \tilde{\bs{\gamma}}_l\tilde{\bs{\gamma}}_l^T. 
 \end{equation}
Subtracting \eqref{eq:part2:4} from  \eqref{eq:part2:3} yields
 \begin{equation} \label{eq:eliminate}
 \sum_{k=1}^K a_{k,1} \bs{\gamma}_k \bs{\gamma}_k^T =  \sum_{l=1}^{\tilde{K}} b_{l,1}  \tilde{\bs{\gamma}}_l \tilde{\bs{\gamma}}_l^T. 
 \end{equation}
For  any $j, j' \in \{1, \dots  p\}$, let $\mb{M}$ satisfy $M_{j,j'} = M_{j',j} = 1$, and all the other entries are zero.  Equations \eqref{eq:part2:2} and \eqref{eq:part2:4} yield
\begin{align} \label{eq:part2:5}
 \sum_{k = 1}^K a_{k,1} \gamma_{k,j} \gamma_{k,j'} \bs{\gamma}_k\bs{\gamma}_k^T  =  \sum_{l = 1}^{\tilde{K}} b_{l,1} \tilde{\gamma}_{l,j} \tilde{\gamma}_{l,j'} \bs{\tilde{\gamma}}_l\bs{\tilde{\gamma}}_l^T. 
\end{align}
Given  \Cref{assumption:1} (c) that $\bs{\Gamma}$ is column full-rank, it can be inferred that there exist $j = j_0$ and $j' = j'_0$ such that  $\gamma_{k, j_0} \gamma_{k, j'_0}\neq 0$ for $k \in \{1, \dots, K\}$.  We prove this by contradiction.  Assume that for all $j, j'$, there exists a $k' \in \{1, \dots, K\}$ such that $\gamma_{k', j} \gamma_{k', j'} = 0$. This implies that $\bs{\gamma}_{k'} = \mb{0}$, contradicting \Cref{assumption:1} (c). Therefore, for  $j = j_0$ and $j'= j'_0$, the rank of the left side of \eqref{eq:part2:5} is $K$, which must match the rank on the right side. Since $\tilde{K} \leq K$ and $\tilde{\bs{\Gamma}}$ is column full-rank, $\tilde{K} = K$.

 Multiplying \eqref{eq:eliminate} by $\gamma_{1,j} \gamma_{1,j'}$ yields 
\begin{align} \label{eq:part2:new:7}
 \sum_{k = 1}^K a_{k,1}  \gamma_{1,j} \gamma_{1,j'}  \bs{\gamma}_k\bs{\gamma}_k^T  =  \sum_{l = 1}^K b_{l,1}   \gamma_{1,j} \gamma_{1,j'} \tilde{\bs{\gamma}}_l\tilde{\bs{\gamma}}_l^T. 
\end{align}
Subtracting \eqref{eq:part2:new:7} from \eqref{eq:part2:5} gives 
\begin{align} \label{eq:part2:4.2}
 \sum_{k = 2}^K a_{k,1} ( \gamma_{k,j} \gamma_{k,j'} - \gamma_{1,j} \gamma_{1,j'})  \bs{\gamma}_k\bs{\gamma}_k^T  =  \sum_{l = 1}^K b_{l,1}  (\tilde{\gamma}_{l,j} \tilde{\gamma}_{l,j'} -  \gamma_{1,j} \gamma_{1,j'})  \tilde{\bs{\gamma}}_l \tilde{\bs{\gamma}}_l^T.  
\end{align}
This cancels out $\bs{\gamma}_1\bs{\gamma}_1^T$ from the left side. Under \Cref{assumption:1} (c), we prove that there exist a $j' = j_1'$ such that $\gamma_{k,j} \gamma_{k,j'} - \gamma_{1,j} \gamma_{1,j'} \neq 0$ for  $j \in \{1, \dots , p\}$ and $k \in \{2, ..., K\}$. By contradiction, if there exist $k^{\ast}$ and $j^{\ast}$ such that $\gamma_{k^{\ast},j^{\ast}} \gamma_{k^{\ast},j'} =  \gamma_{1,j^{\ast}} \gamma_{1,j'}$ for $j' \in \{1, \dots, p\}$,  then $\gamma_{k^{\ast},j'} \propto  \gamma_{1,j'}$ for all $j'$.  Due to $||\bs{\gamma}_{k^{\ast}}||_2 = ||\bs{\gamma}_{1}||_2 = 1$, we have $\bs{\gamma}_{k^{\ast}} = (-1)^l \bs{\gamma}_{1}$ for $l \in \{0, 1\}$, contradicting \Cref{assumption:1} (c). Let $j' = j_1'$ in  \eqref{eq:part2:4.2}, the rank of the left side of the equation is $K-1$, which needs to match the rank on the right side of the equation. Therefore, there must exist a $k' \in \{1, \dots, K\}$ such that 
$\tilde{\gamma}_{k',j} \tilde{\gamma}_{k',j'} - \gamma_{1,j} \gamma_{1,j'}  = 0$ for $j \in \{1, \dots, p\}$, giving  $\tilde{\bs{\gamma}}_{k'} = (-1)^l \bs{\gamma}_1$  for $l \in \{0, 1\}$. Similarly, we apply the same technique to cancel out $\bs{\gamma}_2\bs{\gamma}_2^T, \dots, \bs{\gamma}_K\bs{\gamma}_K^T$ one by one, and establish that  there exists a permutation $\pi(1), \dots, \pi(K)$ of  $\{1,..., K\}$ and $l_j \in \{0, 1\}$ for $j \in \{1, \dots, K\}$, such that $\bs{\gamma}_j = (-1)^{l_j}\tilde{\bs{\gamma}}_{\pi(j)}$.

For the equivalence of the coefficients, without loss of generality, we let $\pi(j) = j$. Similarly to Part (1), we let $\bs{\bs{\gamma}_k^T\mb{M}\bs{\gamma}_k}  = z_k$ for $k \in \{1,..., K\}$. Subtracting  \eqref{eq:part2:4} from  \eqref{eq:part2:2} yields
\begin{align} \label{eq:part2:8}
 \sum_{k = 1}^K a_{k,1} z_k\bs{\gamma}_k\bs{\gamma}_k^T  =  \sum_{l = 1}^{K} b_{l,1} z_l\bs{\gamma}_l\bs{\gamma}_l^T.  
\end{align}
For any $j \in \{1,..., K\}$, setting $z_j = 1$ and $z_j' = 0$ for $j' \neq j$ in \eqref{eq:part2:8} yields $a_{j,1} = b_{j,1}$.  Then we rewrite \eqref{eq:part2:1} as 
\begin{align*}
 \sum_{k = 1}^K a_{k,1} z_k^2  +  \sum_{k = 1}^K a_{k,2} z_k=  \sum_{l = 1}^{K} b_{l,1} z_l^2  + \sum_{l = 1}^K b_{l,2} z_l.
\end{align*} 
For any $j \in \{1,..., K\}$, setting $z_j = 1$ and $z_j' = 0$ for $j' \neq j$ yields $a_{j,1} +a_{j,2}  = b_{j,1} + a_{j,2}$. Since $a_{j,1} = b_{j,1}$, we conclude $a_{j,2} = b_{j,2}$, thereby completing the proof. 

\subsection*{Appendix B: Posterior inference}
We derive the posterior distribution required  to perform posterior sampling. First we derive the marginal conditional posterior distribution in \eqref{eq:posterior:gamma} in 
 \Cref{subsec:sim} following \citet{antoniadis2004bayesian}, which is used to perform Step 1 in  \Cref{alg:pbr}. Then we derive the conditional posterior distributions required for Steps 2 and 3 in \Cref{alg:pbr}.  \\
 
\noindent 
\textbf{Step 1}:  To derive $\int_{\sigma^2} \left(\int_{\mb{c}} p(\mb{r}| \bs{\gamma}, \mb{c}, \sigma^2 )p(\mb{c}|\bs{\gamma}) d \mb{c} \right) p(\sigma^2)  d\sigma^2$, we first write the probability to be integrated 
\begin{align*}
p(\mb{r}| \bs{\gamma}, \mb{c}, \sigma^2) p(\mb{c}|\bs{\gamma}) &\propto \frac{1}{\sigma^n}\text{exp}\left( - \frac{1}{2\sigma^2} (\mb{r} - \tilde{\mb{B}}_{\bs{\gamma}} \mb{c})^T(\mb{r} - \tilde{\mb{B}}_{\bs{\gamma}}\mb{c})\right) \\
& \times \frac{1}{ \sigma^J \text{det}(\bs{\Sigma}_{0, \bs{\gamma}})^{1/2}} \text{exp}\left( -\frac{1}{2\sigma^2} (\mb{c} - \mb{c}_{0, \bs{\gamma}})^T \bs{\Sigma}_{0, \bs{\gamma}}^{-1} (\mb{c} - \mb{c}_{0, \bs{\gamma}}) \right), 
\end{align*}
The two exponential terms above can be written as the product of three exponential components: 
\begin{align*}
&\text{exp}\left( - \frac{1}{2\sigma^2} (\mb{r} - \tilde{\mb{B}}_{\bs{\gamma}} \mb{c})^T(\mb{r} - \tilde{\mb{B}}_{\bs{\gamma}}\mb{c})\right)   \text{exp}\left( -\frac{1}{2\sigma^2} (\mb{c} - \mb{c}_{0, \bs{\gamma}})^T \bs{\Sigma}_{0, \bs{\gamma}}^{-1} (\mb{c} - \mb{c}_{0, \bs{\gamma}}) \right)  \\
& =  \underbrace{\text{exp}\left(-\frac{1}{2\sigma^2} \mb{r}^T\mb{r} \right)}_{F_1(\sigma^2)} \underbrace{\text{exp}\left(-\frac{1}{2\sigma^2} \mb{c}_{0, \bs{\gamma}}^T \bs{\Sigma}_{0, \bs{\gamma}}^{-1} \mb{c}_{0, \bs{\gamma}} \right)}_{F_2(\sigma^2)}  \underbrace{\text{exp}\left(-\frac{1}{2\sigma^2} \left(\mb{c}^T \tilde{\mb{B}}_{\bs{\gamma}}^T\tilde{\mb{B}}_{\bs{\gamma}}  \mb{c}  - 2\mb{r}^T \tilde{\mb{B}}_{\bs{\gamma}} \mb{c}  +  \mb{c}^T \bs{\Sigma}_{0, \bs{\gamma}}^{-1} \mb{c} - 2 \mb{c}^T \bs{\Sigma}_{0, \bs{\gamma}}^{-1}  \mb{c}_{0, \bs{\gamma}}\right) \right)}_{F_3(\mb{c, \sigma^2})},
\end{align*}
where only the third component $F_3(\mb{c}, \sigma^2)$ contains $\mb{c}$, and its integration with respect to $\mb{c}$ yields: 
$$\int_{\mb{c}} F_3(\bs{\mb{c}}, \sigma^2)d\mb{c} = (2\pi)^{J/2} \text{det}\left(\tilde{\bs{\Sigma}}_{\bs{\gamma}}\right)^{1/2} \text{exp}\left(\frac{1}{2}  \tilde{\bs{\mu}}_{\bs{\gamma}}^T \tilde{\bs{\Sigma}}_{\bs{\gamma}}^{-1} \tilde{\bs{\mu}}_{\bs{\gamma}} \right),$$
with
$\tilde{\bs{\Sigma}}_{\bs{\gamma}} = \sigma^2 \left(\tilde{\mb{B}}_{\bs{\gamma}}^T\tilde{\mb{B}}_{\bs{\gamma}} + \bs{\Sigma}_{0,\bs{\gamma}}^{-1}\right)^{-1} = \sigma^2\bs{\Sigma}_{0, \bs{\gamma}}/2$, and
 $\tilde{\bs{\mu}}_{\bs{\gamma}} = \tilde{\bs{\Sigma}}_{\bs{\gamma}} (\tilde{\mb{B}}_{\bs{\gamma}}^T\mb{r} +\bs{\Sigma}_{0, \bs{\gamma}}^{-1} \mb{c}_{0, \bs{\gamma}})/\sigma^2 = (\bs{\Sigma}_{0, \bs{\gamma}} + \bs{\Sigma}_{\rho}) \tilde{\mb{B}}_{\bs{\gamma}}^T\mb{r}/2.$ Therefore, 
 \begin{align*}
 	 \int_{\sigma^2} \left(\int_{\mb{c}} p(\mb{r}| \bs{\gamma}, \mb{c}, \sigma^2 )p(\mb{c}|\bs{\gamma}) d \mb{c} \right) p(\sigma^2)  d\sigma^2 &\propto \int_{\sigma^2}  \frac{1}{ \sigma^{n+J} \text{det}(\bs{\Sigma}_{0, \bs{\gamma}})^{1/2}}  F_1(\sigma^2) F_2(\sigma^2) \text{det}\left(\tilde{\bs{\Sigma}}_{\bs{\gamma}}\right)^{1/2} \text{exp}\left(\frac{1}{2}  \tilde{\bs{\mu}}_{\bs{\gamma}}^T \tilde{\bs{\Sigma}}_{\bs{\gamma}}^{-1} \tilde{\bs{\mu}}_{\bs{\gamma}} \right)p(\sigma^2) d\sigma^2 \\
 	 & \propto \int_{\sigma^2}  \text{exp}\left(- \frac{\mb{r}^T\mb{r}}{2\sigma^2} \right) \text{exp}\left(-\frac{1}{2\sigma^2} \mb{c}_{0, \bs{\gamma}}^T \bs{\Sigma}_{0, \bs{\gamma}}^{-1}\mb{c}_{0, \bs{\gamma}} \right)  \text{exp}\left(\frac{1}{2}  \tilde{\bs{\mu}}_{\bs{\gamma}}^T \tilde{\bs{\Sigma}}_{\bs{\gamma}}^{-1} \tilde{\bs{\mu}}_{\bs{\gamma}} \right) (\sigma^2)^{-\alpha - \frac{n}{2}-1} \text{exp}\left(-\frac{\beta}{\sigma^2}\right) d\sigma^2 \\
 	 &\propto \left(\beta^{\ast}_{\bs{\gamma}} \right)^{-\alpha^{\ast}} \int_{\sigma^2} \phi(\sigma^2 |\alpha^{\ast},\beta^{\ast}_{\bs{\gamma}})d\sigma^2 \\
 	 &= \left(2\beta^{\ast}_{\bs{\gamma}}\right)^{-\alpha^{\ast}},
 \end{align*}
where $\psi(\ \cdot \ | \ \alpha^{\ast},\beta^{\ast}_{\bs{\gamma}})$ is the density of $IG(\alpha^{\ast}, \beta^{\ast}_{\bs{\gamma}})$ with $\alpha^{\ast} = \alpha + n/2$ and $\beta^{\ast}_{\bs{\gamma}} = \mb{r}^T\mb{r}/2 + 
\mb{c}_{0, \bs{\gamma}}^T \bs{\Sigma}_{0, \bs{\gamma}}^{-1}\mb{c}_{0, \bs{\gamma}}/2 + \beta - \tilde{\bs{\mu}}_{\bs{\gamma}}^T \bs{\Sigma}_{0, \bs{\gamma}}^{-1}\tilde{\bs{\mu}}_{\bs{\gamma}}$. We can write 
$2\beta_{\bs{\gamma}}^{\ast} = S(\bs{\gamma}) + 2\beta$, where 
$S(\bs{\gamma}) = \mb{r}^T\mb{r} + 
\mb{c}_{0, \bs{\gamma}}^T \bs{\Sigma}_{0, \bs{\gamma}}^{-1}\mb{c}_{0, \bs{\gamma}} -2 \tilde{\bs{\mu}}_{\bs{\gamma}}^T \bs{\Sigma}_{0, \bs{\gamma}}^{-1}\tilde{\bs{\mu}}_{\bs{\gamma}}$. We substitute 
$\mb{c}_{0, \bs{\gamma}} = \bs{\Sigma}_\rho \tilde{\bs{B}}_{\bs{\gamma}}^T \mb{r}$ and $\tilde{\mu}_{\bs{\gamma}} = (\bs{\Sigma}_{0, \bs{\gamma}} + \bs{\Sigma}_{\rho}) \tilde{\mb{B}}_{\bs{\gamma}}^T\mb{r}/2$ into the expression of  $S(\bs{\gamma})$  and get  $S(\bs{\gamma})= \bs{r}^T \bs{r}  - \bs{r}^T\tilde{\mb{B}}_{\bs{\gamma}} \left( \bs{\Sigma}_{\rho} + \frac{1}{2}\bs{\Sigma}_{0, \bs{\gamma}}  - \frac{1}{2}\bs{\Sigma}_{\rho} \bs{\Sigma}_{0, \bs{\gamma}}^{-1}\bs{\Sigma}_{\rho} \right)  \tilde{\mb{B}}_{\bs{\gamma}}^T\bs{r} $ to complete the derivation. \\

\noindent 
\textbf{Step 2}: Let $\mb{e}_{t-1} = (e_{1,t-1}, \dots, e_{n, t-1})$, where $e_{i,t-1} =  Y_i - \sum_{k=1}^K g_{k,t} \left( \bs{\gamma}_{k,t}^T \mb{M}_i \bs{\gamma}_{k,t} \right)$. 
\begin{align*}
p \left(\sigma^2 |\mu_{t-1}, \{g_{k,t}\}_{k=1}^K, \{\bs{\gamma}_{k,t}\}_{k=1}^K,  \mathcal{D} \right) 
&\propto p \left(\mb{e}_{t-1} \middle | \ \mu_{t-1}, \sigma^2 \right)p(\sigma^2)\\
& \propto  \frac{1}{\sigma^n} \text{exp}\left(- \frac{1}{2 \sigma^2} \sum_{i=1}^n (e_{i,t-1} - \mu_{t-1})^2\right) (\sigma^2)^{-\alpha - 1} \text{exp}\left(- \frac{\beta}{\sigma^2} \right) \\
&\propto (\sigma^2)^{-\frac{n}{2} - \alpha - 1} \text{exp} \left(- \frac{1}{\sigma^2} \left(\beta + \frac{1}{2}\sum_{i=1}^n \left(e_{i,t-1} - \mu_{t-1}\right)^2 \right)\right) 
\end{align*}
Thus, $\sigma^2 |\mu_{t-1}, \{g_{k,t}\}_{k=1}^K, \{\bs{\gamma}_{k,t}\}_{k=1}^K,  \mathcal{D} \sim IG(\tilde{\alpha}, \tilde{\beta})$, where $\tilde{\alpha} = \alpha + n/2$ and $\tilde{\beta} = \beta + \sum_{i=1}^n \left(e_{i,t-1} - \mu_{t-1}\right)^2/2$. \\

\noindent 
\textbf{Step 3}:  
\begin{align*}
p (\mu |\sigma_{t}, \{g_{k,t}\}_{k=1}^K, \{\bs{\gamma}_{k,t}\}_{k=1}^K,  \mathcal{D})  
&\propto p (\mb{e}_{t-1}|\mu, \sigma_t) p(\mu) \\
& \propto  \text{exp}\left(- \frac{1}{2 \sigma_t^2} \sum_{i=1}^n (e_{i, t-1} - \mu)^2\right) \text{exp}\left(-\frac{1}{2 b^2}(\mu - a)^2\right) \\
& \propto  \text{exp}\left(-  \frac{1}{2\sigma_t^2} n\mu^2  - \frac{1}{2b^2} \mu^2   +  \frac{1}{\sigma_t^2} \sum_{i=1}^n e_{i, t-1} \mu  + \frac{a}{b^2} \mu  \right) \\
& \propto  \text{exp}\left(-  \frac{1}{2} \left(\frac{1}{b^2} + \frac{n}{\sigma_t^2}  \right) \mu^2   +  \left(\frac{a}{b^2} + \frac{1}{\sigma_t^2} \sum_{i=1}^n e_{i, t-1} \right) \mu \right). 
\end{align*}
Thus,  $\mu |\sigma^2_{t}, \{g_{k,t}\}_{k=1}^K, \{\bs{\gamma}_{k,t}\}_{k=1}^K,  \mathcal{D} \sim N \left(\tilde{\mu}, \tilde{\sigma}^2 \right),$
where $\tilde{\sigma}^2 = \left( 1/b^2 + n/\sigma_{t}^2\right)^{-1}$ and $\tilde{\mu} = \tilde{\sigma}^2 \left(a/b^2 + \sum_{i=1}^n e_{i, t-1}/\sigma_{t}^2\right).$














\newpage

\bibliographystyle{myjmva.bst}


\bibliography{references.bib}






\end{document}